\documentclass[preprint,12pt]{elsarticle}

\usepackage{graphicx}
\usepackage{amssymb}

\usepackage{lineno}
\usepackage{color}
\usepackage{courier}
\usepackage{tikz}
\usetikzlibrary{calc}
\usetikzlibrary{backgrounds,positioning}
\usetikzlibrary{decorations.pathreplacing}
\usepackage{diagbox}

\newtheorem{theorem}{{\bf Theorem}}

\newtheorem{lemma}{{\bf Lemma}}

\newenvironment{sproof}{\noindent {\bf Proof.}}{\hfill$\Box$}

\usepackage{xcolor}
\usepackage[linesnumbered,ruled,vlined,procnumbered,noresetcount]{algorithm2e}
\usepackage{etoolbox}

\makeatletter
\AtBeginEnvironment{procedure}{\let\c@algocf\c@procedure}
\def\ps@pprintTitle{%
 \let\@oddhead\@empty
 \let\@evenhead\@empty
 \def\@oddfoot{}%
 \let\@evenfoot\@oddfoot}
\makeatother

\usepackage{amsmath}
\DeclareMathOperator*{\argmax}{argmax}
\DeclareMathOperator*{\argmin}{argmin}

\SetCommentSty{mycommfont}

\usepackage{multirow}
\usepackage{booktabs}
\usepackage{multicol}

\usepackage{float}

\usepackage{caption}
\usepackage{subcaption}

\SetKwInput{KwInput}{Input}                
\SetKwInput{KwOutput}{Output}

\SetProcNameSty{texttt}

\tikzset{
	rednode/.style = {rectangle, draw=black!60, fill=red!20, very thick, minimum size=5mm,outer sep=0pt},
	bluenode/.style = {rectangle, draw=black!60, fill=blue!20, very thick, minimum size=5mm,outer sep=0pt},
	greennode/.style = {rectangle, draw=black!60, fill=green!20, very thick, minimum size=5mm,outer sep=0pt},
	whitenode/.style = {rectangle, draw=black!60, fill=white!20, very thick, minimum size=5mm,outer sep=0pt},
}
\usepackage{xcolor}

\usepackage{hyperref} 
\hypersetup{linkbordercolor=white}

\begin{document}

\begin{frontmatter}


\title{A Particle Swarm Inspired Approach for Continuous Distributed Constraint Optimization Problems}





\author[mymainaddress]{Moumita Choudhury}

\author[mymainaddress]{Amit Sarker}

\author[mymainaddress]{Md. Mosaddek Khan}

\author[mysecondaryaddress2]{William Yeoh}

\address[mymainaddress]{Department of Computer Science and Engineering, University of Dhaka}
\address[mysecondaryaddress2]{Department of Computer Science and Engineering, Washington University in St. Louis}




\allowdisplaybreaks
\sloppy

\begin{abstract}
\noindent
\emph{Distributed Constraint Optimization Problems} (DCOPs) are a widely studied framework for coordinating interactions in cooperative multi-agent systems. In classical DCOPs, variables owned by agents are assumed to be discrete. However, in many applications, such as target tracking or sleep scheduling in sensor networks, continuous-valued variables are more suitable than discrete ones. To better model such applications, researchers have proposed \emph{Continuous DCOPs} (C-DCOPs), an extension of DCOPs, that can explicitly model problems with continuous variables. The state-of-the-art approaches for solving C-DCOPs experience either onerous memory or computation overhead and unsuitable for non-differentiable optimization problems. To address this issue, we propose a new C-DCOP algorithm, namely \emph{Particle Swarm Optimization Based C-DCOP} (PCD), which is inspired by \emph{Particle Swarm Optimization} (PSO), a well-known \emph{centralized} population-based approach for solving continuous optimization problems. In recent years, population-based algorithms have gained significant attention in classical DCOPs due to their ability in producing high-quality solutions. Nonetheless, to the best of our knowledge, this class of algorithms has not been utilized to solve C-DCOPs and there has been no work evaluating the potential of PSO in solving classical DCOPs or C-DCOPs. In light of this observation, we adapted PSO, a centralized algorithm, to solve C-DCOPs in a decentralized manner. The resulting PCD algorithm not only produces good-quality solutions but also finds solution without any requirement for derivative calculations. 
Moreover, we design a crossover operator that can be used by PCD to further improve the quality of solutions found. Finally, we theoretically prove that PCD is an anytime algorithm and empirically evaluate PCD against the state-of-the-art C-DCOP algorithms in a wide variety of benchmarks.
\end{abstract}


\end{frontmatter}



\section{Introduction}
\noindent
\emph{Distributed Constraint Optimization Problems} (DCOPs) are an important constraint-handling framework for multi-agent systems in which multiple agents communicate with each other in order to optimize a global objective. The global objective is defined as the aggregation of cost functions (i.e.,~constraints) among the agents. Each of the cost functions involves a set of variables controlled by the corresponding agents. The structure of DCOPs has made it suitable for deploying in various real-world problems. It has been widely applied to solve a number of multi-agent coordination problems including multi-agent task scheduling~\cite{sultanik2007modeling}, sensor networks~\cite{farinelli2014agent}, multi-robot coordination~\cite{Yedidsion2016ApplyingDT}, etc.

Over the years, several algorithms have been proposed to solve DCOPs, and they are broadly categorized into exact and non-exact algorithms. Exact algorithms, such as ADOPT~\cite{modi2005adopt}, DPOP~\cite{Petcu2005ASM, rashik2020speeding}, and PT-FB~\cite{litov2017forward} are designed in such a way that they provide a global optimal solution of a given DCOP. However, since DCOPs are NP-Hard, exact algorithms experience exponential memory requirements and/or exponential computational costs as the system grows. On the contrary, non-exact algorithms such as DSA~\cite{zhang2005distributed}, MGM \& MGM2~\cite{Maheswaran2004Distributed}, Max-Sum~\cite{farinelli2008decentralised,khangdp, khan2018speeding}, CoCoA~\cite{Leeuwen2017CoCoAAN},  ACO\_DCOP~\cite{chen2018ant}, and AED~\cite{mahmud2019aed} compromise some solution quality for scalability.

In general, DCOPs assume that the variables of participating agents are discrete. Nevertheless, many real-world applications (e.g., target tracking sensor orientation~\cite{fitzpatrick2003distributed}, sleep scheduling of wireless sensors~\cite{hsin2004network}) can be best modeled with continuous variables. Therefore, for discrete DCOPs to be applied in such problems, we need to discretize the continuous domains of the variables. However, the discretization process needs to be coarse for a problem to be tractable and must be sufficiently fine to find high-quality solutions of the problem~\cite{stranders2009decentralised}. To overcome this issue, a continuous version of DCOPs have been proposed~\cite{stranders2009decentralised}, which is later referred to as both \emph{Functional DCOPs}~\cite{choudhury2020particle} and \emph{Continuous DCOPs} (C-DCOPs)~\cite{hoang2020new}. In this paper, we will refer to it as C-DCOPs following the most popular convention. There are two main differences between C-DCOPs and DCOPs. Firstly, instead of having discrete decision variables, C-DCOPs have continuous variables that can take any value between a range.  Secondly, the constraint functions are represented in functional forms in C-DCOPs rather than in the tabular forms in DCOPs.

In order to cope with the modification of the DCOP formulation, several C-DCOP algorithms have been proposed. Similar to DCOP algorithms, C-DCOP algorithms are also classified as exact and non-exact approaches (detailed discussions can be found in Section~\ref{section:rel}). In this paper, we focus on the latter class of C-DCOP algorithms as the ensuing exponential growth of search space can make exact algorithms computationally infeasible to deploy in practice. Now, the state-of-the-art algorithms for C-DCOPs are based on either inference~\cite{stranders2009decentralised,voice2010hybrid,hoang2020new} or local search~\cite{hoang2020new}. In the inference-based C-DCOP algorithms, discrete inference-based algorithms, such as Max-Sum and DPOP, have been used in combination with continuous non-linear optimization methods. And, in the only local search-based C-DCOP algorithm, the discrete local search-based algorithm DSA has been extended with continuous optimization methods. However, continuous optimization methods, such as gradient-based optimization require derivative calculations and are thus not suitable for non-differentiable optimization problems.

Against this background, we propose a \emph{Particle Swarm Optimization} (PSO) based C-DCOP algorithm called \emph{PSO-Based C-DCOP} (PCD).\footnote{Preliminary versions of this research have appeared previously~\cite{choudhury2020particle}. This paper contains a more efficient approach and comprehensive description of the algorithm and comes with broader theoretical and experimental analysis to other state-of-the-art C-DCOP algorithms.} PSO is a stochastic optimization technique inspired by the social metaphor of bird flocking~\cite{eberhart1995particle}. It has been successfully applied to many optimization problems such as Function Minimization~\cite{shi1999empirical}, Neural Network Training~\cite{zhang2007hybrid}, and Power-System Stabilizers Design Problems~\cite{abido2002optimal}. However, to the best of our knowledge, no previous work has been done to incorporate PSO in distributed scenarios similar to DCOPs or C-DCOPs. In PCD, agents cooperatively keep a set of particles where each particle represents a candidate solution and iteratively updates the solutions using a series of update equations over time. Since PSO requires only primitive mathematical operators such as addition and multiplication, it is computationally less expensive (both in memory and speed) than the gradient-based optimization methods. Furthermore, PSO is a widely studied technique with a variety of parameter choices and variants developed over the years. Hence, the wide opportunity for developing PCD as a robust population-based algorithm has inspired us to analyze the challenges and opportunities of PSO in C-DCOPs. 
Our main contributions are as follows.
\begin{itemize}
    \item We develop a new algorithm PCD by tailoring PSO. In so doing, we redesign a series of update equations that utilize the communication topology in a distributed scenario.
    \item We introduce a new crossover operator that further improves the quality of solutions found and name the version PCD\_CrossOver.
    \item We analyze the various parameter choices of PCD that balance exploration and exploitation.
    \item We provide a theoretical proof of anytime convergence of our algorithm, and show empirical evaluations of PCD and PCD\_CrossOver on various C-DCOP benchmarks. The results show that the proposed approach finds solutions with better quality by exploring a large search space compared to existing C-DCOP solvers.
\end{itemize}


In Section \ref{section:rel}, we briefly review related work. In Section \ref{section:problem}, we formulate the DCOP and C-DCOP frameworks as well as introduce PSO. Section \ref{section:algo} illustrates the details of our proposed PCD framework. Section \ref{section:theory} provides a theoretical proof of the anytime property and complexity analyses of PCD. In Section \ref{section:exp}, we show empirical evaluations of PCD against existing C-DCOP algorithms. Finally, Section \ref{section:future} concludes the findings of the paper and provides insights for future work.

\section{Related Work}\label{section:rel}
\noindent
In this section, we discuss existing state-of-the-art exact and non-exact C-DCOP algorithms. The only exact algorithm for C-DCOP is the \emph{Exact Continuous DPOP} (EC-DPOP), which only provides exact solutions to linear and quadratic cost functions and is defined over tree-structured graphs only~\cite{hoang2020new}. While there are several non-exact algorithms exist, the first non-exact algorithm for C-DCOP is the \emph{Continuous Max-Sum} (CMS) algorithm. CMS extends the discrete Max-Sum~\cite{stranders2009decentralised} by approximating constraint cost functions as piece-wise linear functions. Subsequently, researchers introduced \emph{Hybrid Continuous Max-Sum} (HCMS), which extends CMS by combining it with continuous non-linear optimization methods~\cite{voice2010hybrid}. However, continuous optimization methods, such as gradient-based optimization require derivative calculations and are thus not suitable for non-differentiable optimization problems. Finally, \citeauthor{hoang2020new}~\cite{hoang2020new} made the most recent contributions to this field. In their paper, the authors proposed four algorithms -- one exact and three non-exact C-DCOP solvers. The exact algorithm is EC-DPOP, which we discussed earlier. The non-exact algorithms are \emph{Approximate Continuous DPOP} (AC-DPOP), \emph{Clustered AC-DPOP} (CAC-DPOP), and \emph{Continuous DSA} (C-DSA). Both AC-DPOP and CAC-DPOP are based on the discrete DPOP algorithm with non-linear optimization techniques. The discrete DPOP algorithm~\cite{Petcu2005ASM} is an inference-based DCOP algorithm that performs dynamic programming on a pseudo-tree representation of the given problem. This algorithm only requires a linear number of messages but has an exponential memory requirement and sends exponentially large message sizes. Since the underlying algorithm for AC-DPOP is DPOP, it also suffers from the same exponentially large message sizes, which is a limiting factor for communication-constrained applications. Although CAC-DPOP provides a bound on the message size by limiting the number of tuples to be sent in the messages, each agent still needs to maintain the original set of tuples in their memory for better accuracy in calculation. Hence, CAC-DPOP still incurs an exponential memory requirement. Nevertheless, the authors also provide C-DSA, a local search algorithm based on DSA. Unlike the DPOP variants, C-DSA's memory requirement is linear in the number of variables of the problem and it sends constant-size messages.

\section{Background and Problem Formulation}
\label{section:problem}
\noindent
In this section, we formulate the problem and discuss the background necessary to understand our proposed method. We first describe the general DCOP framework and then move to the C-DCOP framework, which is our problem of interest in this paper. We then discuss the centralized PSO algorithm and the challenges in incorporating PSO with the C-DCOP framework. 

\subsection{Distributed Constraint Optimization Problems}
\noindent
A \emph{Distributed Constraint Optimization Problem} (DCOP) can be defined as a tuple $ \langle A,X,D,F,\alpha \rangle $~\cite{modi2005adopt} where,
\begin{itemize}
    \item $A$ is a set of agents $\{a_1,a_2,\ldots,a_n\}$.
    \item $X$ is a set of discrete variables $\{x_1,x_2,\ldots,x_m\}$, where each variable $x_j$ is controlled by one of the agents  $a_i$ $\in$ $A$.
    \item $D$ is a set of discrete domains $\{D_1, D_2,\ldots,D_m\}$, where each $D_i$ corresponds to the domain of variable $x_i$.
    \item $F$ is a set of cost functions $\{f_1,f_2,\ldots,f_l\}$, where each $f_i \in F$ is defined over a subset $x^i$ = \{$x_{i_{1}}$, $x_{i_{2}}$, \ldots, $x_{i_{k}}$\} of variables $X$, called the \emph{scope} of the function, and the cost for the function $f_i$ is defined for every possible value assignment of  $x^i$, that is, $f_i$: $D_{i_{1}}$ $\times$ $D_{i_{2}}$ $\times\ldots\times$ $D_{i_{k}}$ $ \to \mathbb{R}$, where the arity of the function $f_i$ is $k$. In this paper, we consider only binary cost functions (i.e.,~there are only two variables in the scope of all functions).
    \item $\alpha: X \rightarrow
	A$ is a variable-to-agent mapping function~\cite{khan2018near} that assigns the control of each variable $x_j \in X$ to an agent $a_i$ $\in$ $A$. Each agent can hold several variables. However,  for the ease of understanding, we assume each agent controls only one variable in this paper.
\end{itemize}
An optimal solution of a DCOP is an assignment $X^*$ that minimizes the sum of cost functions as shown in Equation \ref{eq:1}\footnote{For a maximization problem, the $\argmin$ operator should be replaced by the $\argmax$ operator.}: 

\begin{align}
    X^* = \argmin_X \sum_{f_i \in F} f_i(x^i)
\label{eq:1}
\end{align}

\subsection{Continuous Distributed Constraint Optimization Problems}
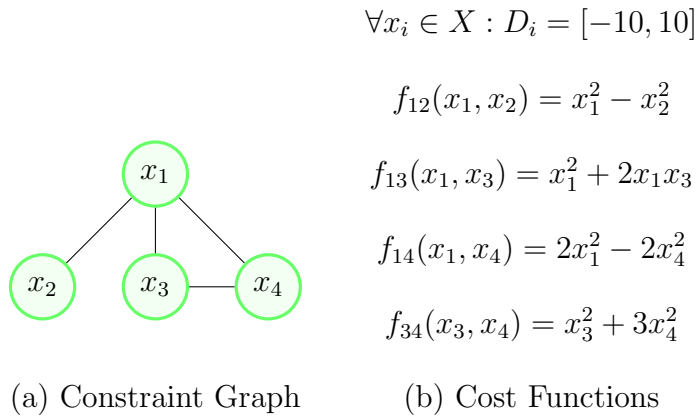
\begin{figure}[t]
\centering
  \begin{tikzpicture}
        [
        roundnode/.style={circle, draw=green!60, fill=green!5, very thick, minimum size=7mm},
        ]
        \node[roundnode]    at(0.5,0)  (x1)                              {$x_1$};
        \node[roundnode]    at(-1,-1.5)  (x2)                              {$x_2$};
        \node[roundnode]    at(0.5,-1.5)  (x3)                              {$x_3$};
        \node[roundnode]    at(2,-1.5)  (x4)                              {$x_4$};
         
        \draw (x1) -- (x2);
        \draw (x1) -- (x3);
        \draw (x3) -- (x4);
        \draw (x4) -- (x1);
        \node  at (0.5,-3)
        {
            (a) Constraint Graph
        };
        \node at (5.5,1)  {
           $f_{12}(x_1, x_2) = x_{1}^{2} - x_{2}^{2}$
       
        };
        \node at (5.5,0)  {
             $f_{13}(x_1, x_3) = x_{1}^{2} + 2x_{1}x_{3}$
        };
        \node at (5.5,-1)  {
           $f_{14}(x_1, x_4) = 2x_{1}^{2} - 2x_{4}^{2}$
        };
        \node at (5.5,-2)  {
            $f_{34}(x_3, x_4) = x_{3}^{2} + 3x_{4}^{2}$
        };
        \node at (5.5,2)  {
             $\forall x_i \in X: D_i = [-10, 10]$
        };
        \node  at (5.5,-3)
        {
            (b) Cost Functions 
        };
\end{tikzpicture}
\caption{Example of a C-DCOP.}
\label{fdcopex}
\end{figure}
\noindent
Similar to the DCOP formulation, C-DCOPs can be defined as a tuple $ \langle A,X,D,F,\alpha \rangle$~\cite{hoang2020new}. In C-DCOPs, $A$, $F$, and $\alpha$ are the same as defined in DCOPs. Nonetheless, the set of variables $X$ and the set of domains $D$ are defined as follows:
\begin{itemize}
    \item $X$ is the set of \emph{continuous} variables $\{x_1,x_2,\ldots,x_m\}$, where each variable $x_j$ is controlled by one of the agents $a_i \in A$.
    \item $D$ is a set of \emph{continuous} domains $\{D_1, D_2,\ldots,D_m\}$, where each $D_i = [LB_i, UB_i]$ corresponds to the domain of variable $x_i$. In other words, variable $x_i$ can take on any value in the range of $LB_i$ to $UB_i$.
\end{itemize}
As discussed in the previous section, a notable difference between DCOPs and C-DCOPs can be found in the representation of the cost functions. In DCOPs, the cost functions are conventionally represented in the form of a table, while in C-DCOPs, they are represented in the form a function~\cite{hoang2020new}. However, the goal of a C-DCOP remains the same as depicted in Equation \ref{eq:1}. Figure~\ref{fdcopex} presents an example C-DCOP, where Figure~\ref{fdcopex}a shows a constraint graph with four variables with each variable $x_i$ controlled by an agent $a_i$. Each edge in the constraint graph represents a cost function and the definition of each function is shown in Figure~\ref{fdcopex}b. In this particular example, the domains of all variables are the same -- each variable $x_i$ can take values from the range $[-10, 10]$. 

\subsection{Particle Swarm Optimization}
\noindent
\emph{Particle Swarm Optimization} (PSO) is a population-based optimization\footnote{For simplicity, we are going to consider the terms `optimization' and `minimization' interchangeably throughout the paper.} technique inspired by the movement of a bird flock or a fish school~\cite{eberhart1995particle}. In PSO, each individual of the population is called a particle. PSO solves the problem by moving the particles in a multi-dimensional search space by adjusting the position and velocity of each particle. As shown in Algorithm~\ref{algo:pso}, each particle is initially assigned a random position and velocity (Line~\ref{line:2}). A fitness function is defined, which is used to evaluate the position of each particle. In each iteration, the movement of a particle is guided by both its \emph{local best position} found so far in the search space and the \emph{global best position} found by the entire swarm (Lines~\ref{line:5}-\ref{line:7}). The combination of the local and global best positions ensures that when a global better position is found through the search process, the particles will move closer to that position and explore the surrounding search space more thoroughly. Then, the local best position of each particle and the global best position of the entire population is updated when necessary (Lines~\ref{line:8}-\ref{line:11}). Over the last couple of decades, several versions of PSO have been developed. The standard PSO often converges to a sub-optimal solution since the velocity component of the global best particle tends to zero after some iterations. Consequently, the global best position stops moving, and the swarm behavior of all other particles leads them to follow the global best particle. To cope with the premature convergence property of standard PSO, Guaranteed Convergence PSO (GCPSO) has been proposed that provides convergence guarantees to a local optima~\cite{van2002new}.
\begin{algorithm}[t]
\DontPrintSemicolon

Generate an $n$-dimensional population $P$\;\label{line:1}
   Randomly initialize positions and velocities of each particle\;\label{line:2}
   \While{termination condition is not met \label{line:3}}
   {
   		\ForEach{$P_i \in P$ \label{line:4}}{
   		    calculate the current velocity \;\label{line:5}
   		    calculate the next position given current velocity \; \label{line:6}
   		    move to next position\;\label{line:7}
   		    \If{fitness of current position $<$ fitness of local best \label{line:8}}{
   		    update local best\;\label{line:9}
   		    }
   		    \If{fitness of current position $<$ fitness of global best \label{line:10}}{
   		    update global best\;\label{line:11}
   		    }
   		    
   		}
   }
\caption{Particle Swarm Optimization}
\label{algo:pso}
\end{algorithm}

\subsection{Challenges}
\noindent
Over the years, PSO and its improved variant \emph{Guaranteed Convergence PSO} (GCPSO) have shown promising performance in centralized continuous optimization problems~\cite{shi1999empirical, van2002new}. Motivated by its success, we seek to explore its potential in solving C-DCOPs. However, there are several challenges that must be addressed when developing an anytime C-DCOP algorithm using GCPSO:
\begin{itemize}
\item \textbf{Particles and Fitness Representation:}
We need to define a representation for the particles where each particle represents a solution of the C-DCOPs.  Moreover, a distributed method for calculating the fitness for each of the particles needs to be devised. 
\item \textbf{Creating the Population:} In centralized optimization problems, creating the initial population is a trivial task. However, in the case of C-DCOPs, different agents control different variables. Hence, a method needs to be devised to generate the initial population cooperatively.
\item \textbf{Evaluation:} Centralized PSO deals with an $n$-dimensional optimization task. In C-DCOPs, each agent holds one variable and each agent is responsible for solving the optimization task related to that variable only where the global objective is still an $n$-dimensional optimization process. Thus, a decentralized evaluation needs to be devised.
\item \textbf{Maintaining the Anytime Property:} To maintain the anytime property in a C-DCOP approach, we need to identify the global best particle and the local best position for each particle. A distribution method needs to be devised to notify all the agents when a new global best particle or local best position is found. Finally, a decentralized coordination method is needed among the agents to update the position and velocity considering the current best position.
\end{itemize}
In the following section, we devise a novel method that addresses the above challenges and applies PSO to solve C-DCOPs.

\begin{figure}[t]
\centering

  \begin{tikzpicture}
        [
        roundnode/.style={circle, draw=green!60, fill=green!5, very thick, minimum size=7mm},
        ]
        
        \node[roundnode]    at(0.5,0)  (x1)                              {$x_1$};
        \node[roundnode]    at(-1,-1.5)  (x2)                              {$x_2$};
        \node[roundnode]    at(0.5,-1.5)  (x3)                              {$x_3$};
        \node[roundnode]    at(2,-1.5)  (x4)                              {$x_4$};
         
        \draw (x1) -- (x2);
        \draw (x1) -- (x3);
        \draw[dotted] (x3) -- (x4);
        \draw (x4) -- (x1);
        
\end{tikzpicture}
\caption{A sample BFS pseudo-tree representation of the C-DCOP depicted in Figure~\ref{fdcopex}.}
\label{bfsptree}
\end{figure}
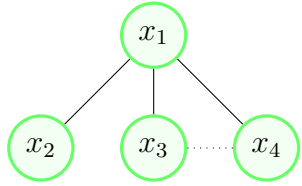

\section{The PCD Algorithm}\label{section:algo}
\begin{table}[]
\centering
\caption[Population Representation in PCD ]{%
  Population Representation in PCD }
\begin{tabular}{@{}lllll@{}}
\toprule
                                & Agent $a_1$ & Agent $a_2$ & Agent $a_3$               & Agent $a_m$ \\ \midrule
\multirow{2}{*}{Particle $P_1$} & $P^1_1.v_1$ & $P^2_1.v_2$ & \multirow{2}{*}{$\ldots$} & $P^m_1.v_m$ \\
                                & $P^1_1.x_1$ & $P^2_1.x_2$ &                           & $P^m_1.x_m$ \\ \midrule
\multirow{2}{*}{Particle $P_2$} & $P^1_2.v_1$ & $P^2_2.v_2$ & \multirow{2}{*}{$\dots$}  & $P^m_2.v_m$ \\
                                & $P^1_2.x_1$ & $P^2_2.x_2$ &                           & $P^m_2.x_m$ \\ \midrule
\multirow{2}{*}{$\ldots$} & \multirow{2}{*}{$\ldots$} & \multirow{2}{*}{$\ldots$} & \multirow{2}{*}{$\ldots$} & \multirow{2}{*}{$\ldots$} \\
                                &             &             &                           &             \\ \midrule
\multirow{2}{*}{Particle $P_K$} & $P^1_K.v_1$ & $P^2_K.v_2$ & \multirow{2}{*}{$\ldots$} & $P^m_K.v_m$ \\
                                & $P^1_K.x_1$ & $P^2_K.x_2$ &                           & $P^m_K.x_m$ \\ \bottomrule
\end{tabular}
\label{tab:population}
\end{table}
\noindent
We now turn to describe our proposed \emph{Particle Swarm Optimization Based C-DCOP} (PCD) algorithm. To facilitate an easier understanding of the algorithm, we first describe what each particle represents in the context of C-DCOPs. Like in PSO, each particle in PCD has two attributes -- position and velocity. The position of a particle corresponds to a value assignment to all variables in the C-DCOP. In other words, it is a solution to a given C-DCOP. Moreover, each agent also maintains the local best position of the particle. The velocity of a particle defines the step size that a particle takes in each iteration to change its position and is influenced by the combination of the direction of its local best and global best position. However, unlike in PSO, where a centralized entity controls all particles, each particle in PCD is controlled in a decentralized manner by all deployed agents. Specifically, for each particle, each agent controls only the position and velocity corresponding to its variable. 

In PCD, we define population $P$ as a set of particles that are collectively maintained by all the agents and local population $P^i \subseteq P$ as the subset of the population maintained by an agent $a_i$. For further clarification, we present an example of a population in Table~\ref{tab:population}. Here, each row represents a particle $P_k = \{P^1_k, P^2_k, \ldots, P^m_k\}$, which is the $k^{th}$ solution of the problem. Each column represents an agent $a_i$ and the corresponding attributes that it holds for each particle. For example, in the table, each agent $a_i$ holds two attributes, namely the position attribute $P^i_k.x_i$ and the velocity attribute $P^i_k.v_i$, for each particle $P^i_k \in P^i$. 
Additionally, we use the following notations:
\begin{itemize}
    \item $P_k.X = \{P_k^1.x_1, P_k^2.x_2, \ldots , P_k^m.x_m\}$ and $P_k.V = \{P_k^1.v_1, P_k^2.v_2, \ldots , P_k^m.v_m\}$ to represent the complete position and velocity assignment for each particle $P_k$, respectively.
    \item $P^i.x_i = \{ P^i_1.x_i, P^i_2.x_i, \ldots, P^i_K.x_i\}$ and $P^i.v_i = \{P^i_1.v_i, P^i_2.v_i, \ldots, P^i_K.v_i\}$ to represent the position and velocity assignments of each agent $a_i$ for all the particles, respectively.
    \item $P^i_k.local\_fitness$ to represent the fitness of particle $P^i_k$, that is, the aggregated cost of constraints associated with the neighbors of agent $a_i$.
    \item $P_k.fitness$ and $P^i_k.fitness$ to represent the complete fitness and the fitness that agent $a_i$ calculates for each particle $P_k \in P$ and $P^i_k \in P^i$, respectively.
    \item $P^i.fitness \leftarrow \{P^i_1.fitness, P^i_2.fitness, \ldots, P^i_K.fitness\}$ to represent the set of $P^i_k.fitness$ for all the particles.
    \item  $P^i_k.p_{best}$ and $P^i_k.p_{best}.fitness$ to represent the best position of particle $P^i_k$ thus far and the fitness value of that position, respectively.
    \item $P^*$ to represent the global best particle among all particles. 
    \item $P^i.g_{best}$ and $P^i.g_{best}.fitness$ to represent the position attribute of the global best particle $P^*$ and the fitness value of that position for each agent $a_i$, respectively.
\end{itemize}

PCD is a PSO-based iterative algorithm that first constructs a \emph{Breadth First Search} (BFS) pseudo-tree~\cite{chen2017improved}, which orders the agents, in a pre-processing step. Figure~\ref{bfsptree} illustrates a BFS pseudo-tree constructed from the constraint graph shown in
Figure~\ref{fdcopex} having $x_1$\footnote{We use $a_i$ and $x_i$ interchangeably throughout the paper since each agent controls exactly one variable.} as the root. From this point, we use the notation $N_i$ to refer to the neighboring agents of agent $a_i$ in the constraint graph and the notations $PR_i$ and $CH_i \subseteq N_i$ to refer to the parent agent and set of children agents of agent $a_i$ in the pseudo-tree, respectively.
For example, for agent $x_3$ of Figure~\ref{bfsptree}b, $N_3 = \{x_1, x_4\}$, $PR_3 = x_1$, and $CH_3 = \emptyset$.

\begin{algorithm}[t!]
\small
\DontPrintSemicolon 
\SetKwFunction{FVariableUpdate}{VARIABLE\_UPDATE}
\SetKwFunction{FInit}{INITIALIZATION}
\SetKwInOut{Input}{Input}
\SetKwInOut{Output}{Output}
\SetKwFunction{FBestUpdate}{BEST\_UPDATE}
\SetKwFunction{FCrossover}{CROSSOVER}
\SetKwFunction{FEvaluation}{EVALUATION}

   	\Input{
   	$K$ -- Number of particles\\
   	$w$ -- Inertia weight\\
   	$c_1$ -- Cognitive constant\\
   	$c_2$ -- Social constant\\
   	$max_{s_{c}}$ -- Threshold for success count\\
   	$max_{f_{c}}$ -- Threshold for failure count }
   
   \ForEach{$a_i \in A$ \label{line:13}}{
   		\FInit{}\; \label{line:14}
   	    
        \While{Termination condition not met} 
        {

        $P^i.fitness \leftarrow$ \FEvaluation{} \;\label{line:16}
          
        $P^* \leftarrow $ \FBestUpdate{$P^i.fitness$} \;\label{line:17} 

       	$t \leftarrow t + 1$ \;\label{line:18}
        \FVariableUpdate{$P^*$} \;\label{line:19} 
       
     }
   }
\caption{PCD Algorithm}
\label{algo:simplePCD}
\end{algorithm}

\begin{procedure}[t]
  \caption{INITIALIZATION()( ) \label{proc:init}}
  \DontPrintSemicolon
  $P^i \leftarrow$ set of $K$ particles\;\label{line:20}
  $t \leftarrow s_c \leftarrow f_c \leftarrow 0$ \;
  $\rho \leftarrow 1$ \;\label{line:22}
  \ForEach{$P_k^i \in P^i$}{
  		    $P_k^i.v_i \leftarrow$ 0\;\label{line:24}
  		    $P_k^i.x_i \leftarrow$ a random value from $D_i$ \;\label{line:25}
  		    $P_k^i.p_{best} \leftarrow $ null \;\label{line:26}
  		    $P_k^i.p_{best}.fitness \leftarrow \infty $ \;\label{line:27}
  		}
  		$P^i.g_{best} \leftarrow $ null \;\label{line:28}
  		$P^i.g_{best}.fitness \leftarrow \infty$ \;\label{line:29}
  		Send VALUE($P^i.x_i$) to each agent $a_j \in N_i$\;\label{line:30}
        
\end{procedure}

\begin{procedure}[t]
  \caption{EVALUATION()( ) \label{proc:evaluation}}
  \DontPrintSemicolon
Wait until VALUE($P^j.x_j$) is received from each agent  $a_j \in N_i$\;\label{line:31}
   	         \ForEach{$P_k^i \in P^i$ }{
					$P_k^i.local\_fitness \leftarrow \sum_{a_j \in N_i} ^{} f_{ij}(P_k^i.x_i,P_k^j.x_j) $ \;\label{line:33}
             }
          
            Wait until COST($P^j.fitness$)  is received from each agent $a_j \in CH_i$\;\label{line:34}

            \ForEach{$P_k^i \in P^i$ }{
                $P^i_k.fitness \leftarrow P_k^i.local\_fitness + \sum_{a_j \in CH_i}^{} P_k^j.fitness $ \;\label{line:36}
            }
            
            \eIf{$a_i = $ root}{
                \ForEach{$P_k^i \in P^i$ }{
                    $P^i_k.fitness \leftarrow P_k^i.fitness / 2 $ \;\label{line:39}
                }
            }
            {
               Send COST($P^i.fitness$) to $PR_i$\;\label{line:41}
            }
        \Return $P^i.fitness$\;
        
\end{procedure}

\begin{procedure}[t]
\caption{BEST-UPDATE($P^i.fitness$) \label{proc:bestupdate}}
    \DontPrintSemicolon
    $PB \leftarrow \emptyset$ \;\label{line:43}
    $P^* \leftarrow \emptyset$ \;\label{line:44}
    \eIf{$a_i = $ root \label{line:45}}{
        \ForEach{$P_k^i \in P^i$ \label{line:46}}{
            \If{$P_k^i.fitness < P_k^i.p_{best}.fitness$ \label{line:47}}{
                $P_k^i.p_{best} \leftarrow P_k^i.x_i$\;\label{line:48}
                $P_k^i.p_{best}.fitness \leftarrow P_k^i.fitness$ \;\label{line:49}
                $PB \leftarrow  PB \cup \{P_k^i\}$ \;\label{line:50}
            }
            
            \If{$P_k^i.fitness < P^i.g_{best}.fitness$ \label{line:51}}{
                $P^i.g_{best} \leftarrow P_k^i.x_i$\;\label{line:52}
                $P^i.g_{best}.fitness \leftarrow P_k^i.fitness$ \;\label{line:53}
                $P^* \leftarrow P_k^i$\; \label{line:54}
            }
        }
    }
    {
        Wait until BEST$(PB, P^*)$ is received from $PR_i$\;\label{line:56}
    
         \ForEach{$P_k \in PB$ \label{line:57}}{
            $P_k^i.p_{best} \leftarrow P_k.x_i$\; \label{line:58}
        }
    
        \If{$P^* \neq \emptyset$ \label{line:59}}{
            $P^i.g_{best} \leftarrow P^*.x_i$ \;\label{line:60}
        }
    }
    
    Send BEST$(PB, P^*)$ to each agent $a_j \in CH_i$\;\label{line:61}
    \Return $P^*$ \;\label{line:62}
\end{procedure}

\begin{procedure}[t]
\caption{VARIABLE-UPDATE($P^*$) \label{proc:varupdate}}
    \DontPrintSemicolon
    Calculate $s_c$ and $f_c$ using Equations \ref{eq:7} and \ref{eq:8}\;\label{line:63}
    \ForEach{$P_k^i \in P^i$}{
        \eIf{$P_k^i = P^*$ \label{line:65}}{
            Calculate $P_k^i.v_i$ and $P_k^i.x_i$ using Equations \ref{eq:4} and \ref{eq:5}\;\label{line:66}
        }
        {
            Calculate $P_k^i.v_i$ and $P_k^i.x_i$ using Equations \ref{eq:3} and \ref{eq:5}\;\label{line:68}
        }
    }
    Send VALUE($P^i.x_i$) to each agent $a_j \in N_i$\;\label{line:69}

\end{procedure}

The pseudocode of our PCD algorithm can be found in Algorithm~\ref{algo:simplePCD}. After constructing the pseudo-tree, it runs the following three phases: 
\begin{itemize}
\item \textbf{Initialization Phase:} The agents create an initial population of $K$ particles and initialize their parameters. 
\item \textbf{Evaluation Phase:} The agents calculate the fitness value for each particle in a distributed way. 
\item \textbf{Update Phase:} Each agent keeps track of the best solution found so far, propagates this information to the other agents, and updates its value assignment according to that information. 
\end{itemize}
The agents repeat these last two phases in a loop until some termination condition is met.

We now describe these phases in more detail. In the \textbf{initialization phase}, each agent $a_i \in A$ executes the \texttt{INITIALIZATION} procedure (Procedure \ref{proc:init}), which consists of the following: It first creates a set of $K$ particles $P^i$ and initializes the cycle counter $t$ as well as three other variables $s_c$, $f_c$, and $\rho$ that are used to update the velocity of the particles (Lines~\ref{line:20}-\ref{line:22}).
It then initializes the velocity $P_k^i.v_i$ and position $P_k^i.x_i$ of each particle $P_k^i \in P^i$ to 0 and a random value in $D_i$, respectively (Lines~\ref{line:24}-\ref{line:25}). 
This initialization is aimed at distributing the initial positions of the particles randomly throughout the search space.
It then initializes the best position $P_k^i.p_{best}$ and the corresponding fitness value $P_k^i.p_{best}.fitness$ of each particle $P_k^i \in P^i$ to null and infinity, respectively, since the position has not been evaluated yet (Lines~\ref{line:26}-\ref{line:27}).
Similarly, it initializes the best global position $P^i.g_{best}$ and the corresponding fitness value $P^i.g_{best}.fitness$ to null and infinity, respectively, as well (Lines~\ref{line:28}-\ref{line:29}).
Finally, it sends its position assignments for all particles $P^i.x_i$ in a VALUE message to each of its neighboring agents (Line~\ref{line:30}).

Next, in the \textbf{evaluation phase}, the agents collectively calculate the complete fitness $P_k.fitness$ of each particle $P_k$ using the fitness function shown in Equation \ref{eq:2} in the \texttt{EVALUATION} procedure (Procedure \ref{proc:evaluation}).

\begin{align}
     P_k.fitness &= \frac{1}{2} \sum_{a_i \in A} \sum_{f_j \in F^i} f_j(P_k.x^j)
     \label{eq:2} \\
    F^i &= \{f_j \in F \mid x^j = \{x_i,x_k\} \}     
    \label{eq:Fi}
\end{align}
\noindent Here, $F^i$ is the set of constraints whose scope $x^i$ includes $a_i$ (see Equation~\ref{eq:Fi}) and $P_k.x^j$ is the value assignment of the set of variables in the scope $x^j$ of function $f_j$ for each particle $P_k$.
Note that a single agent cannot calculate the complete fitness value. Instead, it is calculated in a decentralized way by all the agents and then accumulated up the BFS tree towards the root. Specifically, each agent $a_i$ is in charge of computing only $\sum_{f_j \in F^i} f_j(P_k.x^j)$ for each particle $P_k$. 
Further, note that the cost of each function $f_j$ is summed up twice by the two agents in its scope.\footnote{Recall that we consider only binary cost functions in this paper.} Therefore, the complete fitness value is divided by two. 

To calculate the complete fitness value in a decentralized way, each agent first waits for VALUE messages from its neighboring agents (Line~\ref{line:31}). Upon receiving all the VALUE messages, it calculates the costs of all its functions $f^j \in F^i$ and aggregates them in local fitness values $P_k^i.local\_fitness$ (Line~\ref{line:33}). If the agent does not have any children agent, then it assigns $P_k^i.local\_fitness$ to $P_k^i.fitness$ for all particles $P_k^i$ (Line~\ref{line:36}) and sends the set of fitness values of all particles in a COST message to its parent agent (Line~\ref{line:41}). 
If an agent does have children agents, then it waits for COST messages from all its children agents (Line~\ref{line:34}). After receiving the fitness values from all its children, it aggregates the fitness values received with its own local fitness values (Line~\ref{line:36}) and sends the set of aggregated fitness values of all particles in a COST message to its parent agent (Line~\ref{line:41}).

This process repeats until the root agent receives all COST messages from all its children agents and calculates the aggregated fitness values. At this point, note that the cost of each constraint is doubly counted in the aggregated fitness values because the local fitness values of both agents in the scope of the constraint are aggregated together. Thus, the root agent divides the aggregated fitness values by two (Line~\ref{line:39}) before starting the next phase.

Finally, in the \textbf{update phase}, the agents synchronize on their best local and global particles in the \texttt{BEST\_UPDATE} procedure (Procedure \ref{proc:bestupdate}) and update the positions and velocities of their particles in the \texttt{VARIABLE\_UPDATE} procedure (Procedure \ref{proc:varupdate}).

To synchronize their best local and global particles, the root agent first checks if a better local position has been found for each particle (Lines~\ref{line:45}-\ref{line:47}). If this is the case, it updates the best position $P_k^i.p_{best}$ and its corresponding fitness value $P_k^i.p_{best}.fitness$ before storing that particle in the set $PB$ (Lines~\ref{line:48}-\ref{line:50}). The root agent also checks if a better global position has been found (Line~\ref{line:51}). If so, it updates the best global position $P^i.g_{best}$ and its corresponding fitness value $P^i.g_{best}.fitness$ before storing that particle in a variable $P^*$ (Lines~\ref{line:52}-\ref{line:54}). The root agent then sends both $PB$ and $P^*$ in a BEST message to each of its children agents (Line~\ref{line:61}).

When each of its children agents receives the BEST message from the root agent, it iterates over all the particles $P_k$ in the set $PB$, and assigns the positions of those particles as best positions $P_k^i.p_{best}$ of the corresponding particles $P_k^i$ in its local copy (Lines~\ref{line:57}-\ref{line:58}). Similarly, if a better global particle has been found, it assigns the position of that particle $P^*.x_i$ as its best global position $P^i.g_{best}$ (Lines~\ref{line:59}-\ref{line:60}). It then propagates both $PB$ and $P^*$ that it received in the BEST message to each of its children agents (Line~\ref{line:61}). This process repeats down the pseudo-tree until all agents synchronize their best local and global particles. Finally, the agents increment their cycle counters by one (Line~\ref{line:18}).

To update the positions and velocities of the particles, we adapt the update equations used by Guaranteed Convergence PSO (GCPSO)~\cite{van2002new}. At a high level, each agent $a_i$ uses Equations~\ref{eq:4} and~\ref{eq:3} to update the velocities of the global best particle $P^*$ and other particles $P_k^i \in P^i \setminus \{P^*\}$, respectively, and use Equation~\ref{eq:5} to update the positions of all particles $P_k^i \in P^i$ (Lines~\ref{line:65}-\ref{line:68}): 

\begin{align} 
P^*.v_i^{(t)} & = -P^*.x_i^{(t-1)} + P^i.g_{best}^{(t-1)} + w P^*.v_i^{(t-1)} + \rho^{(t)}(1-2r_2)
\label{eq:4} \\
P_k^i.v_i^{(t)} &= w P_k^i.v_i^{(t-1)} +  r_1 c_1  (P_k^i.p_{best}^{(t-1)} - P_k^i.x_i^{(t-1)}) 
 + r_2  c_2  (P^i.g_{best}^{(t-1)} - P_k^i.x_i^{(t-1)})
\label{eq:3} \\
P^i_k.x_i^{(t)} &= P^i_k.x_i^{(t-1)} + P^i_k.v_i^{(t)}
\label{eq:5}
\end{align}

\noindent In these equations, the superscripts $(t)$ denote the value of the variables at the $t^{\text{th}}$ cycle. Here, $w$, $c_1$, and $c_2$ are user-defined input parameters to the algorithm; $r_1$ and $r_2$ are two random values that are uniformly sampled from the range $[0, 1]$ by each agent in each cycle; and $\rho^{(t)}$ is defined using Equation~\ref{eq:6}:

\begin{align}
\rho^{(t)} = 
 \begin{cases} 
      2 \rho^{(t-1)} & \text{if } s_c^{(t-1)} > max_{s_{c}} \\
      0.5 \rho^{(t-1)} & \text{else if } f_c^{(t-1)} > max_{f_{c}} \\
      \rho^{(t-1)} & \text{otherwise} 
   \end{cases}
\label{eq:6}
\end{align}

\noindent where $max_{s_{c}}$ and $max_{f_{c}}$ are user-defined input parameters of the algorithm; and both $s_c$ and $f_c$ are calculated using Equations~\ref{eq:7} and~\ref{eq:8}, respectively:

\begin{align}
    s_c^{(t)} &= 
    \begin{cases} 
      s_c^{(t-1)} + 1 & \text{if } P^{*(t)} \neq P^{*(t-1)}  \\
      0 & \text{otherwise}
  \end{cases}
\label{eq:7} \\
    f_c^{(t)} &= 
    \begin{cases} 
      0 & \text{if } P^{*(t)} \neq P^{*(t-1)}  \\
      f_c^{(t-1)} + 1 & \text{otherwise} 
  \end{cases}
\label{eq:8}
\end{align}

Intuitively, $w$ represents an inertia weight that defines the influence of the velocity of the previous cycle on the velocity in the current cycle. The constants $c_1$ and $c_2$ are called the cognitive and social constants, respectively, in the literature because they affect the terms $P_k^i.p_{best}^{(t-1)} - P_k^i.x_i^{(t-1)}$ and $P^i.g_{best}^{(t-1)} - P_k^i.x_i^{(t-1)}$, which are called cognition and social components, respectively. The cognition component is called such because it considers the particle's own attributes only while the social component is called such because it involves interactions between two particles. Both of the constants $c_1$ and $c_2$ define the influence of local and global best positions on the velocity of particles in the current cycle.

The parameter $\rho$ represents the diameter of an area around the global best particle that particles can explore. Its value is determined by the count of consecutive successes $s_c$ and failures $f_c$. Success is defined when the fitness value of the global best particle improves, and failure is defined when the fitness value remains unchanged. When there are more consecutive successes than a threshold $max_{s_{c}}$, the diameter $\rho$ doubles to increase random exploration because the current location of the best particle is promising. On the other hand, when there are more consecutive failures than a threshold $max_{f_{c}}$, the diameter $\rho$ is halved to focus the search closer around the location of the best particle.  

\subsection{Crossover}
\label{sec:crossover}
\begin{procedure}[t]
\caption{CROSSOVER() \label{proc:crossover}}
  \DontPrintSemicolon
   Calculate $P_k^i.b_p$ using Equation \ref{eq:bp}\;\label{line:70}
   Choose two particle $P_a^i$ and $P_b^i$.\;\label{line:71}
   Calculate $P_a^i.v_i$ and $P_a^i.x_i$ using Equations \ref{eq:cross_va} and \ref{eq:cross_xa}\;\label{line:72}
   Calculate $P_b^i.v_i$ and $P_b^i.x_i$ using Equations \ref{eq:cross_vb} and \ref{eq:cross_xb}\;\label{line:73}
\end{procedure}
\noindent
Although PCD provides reasonable anytime solution quality in several benchmark problems (see details in Section \ref{sec:exp_performance}), the scope for incorporating other genetic operators still exists. Hence, in this section, we introduce a new crossover operator that further improves the solution quality of PCD. We refer to this version of PCD with the new crossover operator as PCD\_CrossOver. In centralized hybrid PSO models, arithmetic crossover of position and velocity vectors have shown promising results~\cite{lovbjerg2001hybrid}. In a centralized scenario, algorithms can execute crossover operations simultaneously for all the variables. But in a distributed scenario, either the agents need to agree in a cooperative crossover execution~\cite{chen2020genetic} and need to exchange information or the agents can execute crossover operation only for the variables that they hold. In this paper, we follow the latter approach to not incur additional messaging and synchronization overheads. We describe the crossover operation in Procedure~\ref{proc:crossover}.
Specifically, each agent $a_i$ uses the local fitness value $P_k^i.local\_fitness$ of each particle $P_k^i$ in the evaluation phase (Line~\ref{line:33}) to calculate the crossover probability $P_k^i.b_p$ for each particle $P_k^i$ using Equation~\ref{eq:bp} (Line~\ref{line:70}). 

\begin{align}
    P_k^i.b_p &= \frac{| P_k^i.local\_fitness |}{\sum_{j=1}^{K} | P_j^i.local\_fitness |} 
\label{eq:bp}
\end{align}

Then, each agent $a_i$ selects two random particle $P_a^i$ and $P_b^i$ from its set $P^i$ according to the crossover probabilities (Line~\ref{line:71}), and updates their positions using the following crossover operations (Lines~\ref{line:72}-\ref{line:73}): 

\begin{align}
P_a^i.x_i^{(t)} &= r P_a^i.x_i^{(t-1)} + (1 - r) P_b^i.x_i^{(t-1)} \label{eq:cross_xa} \\
P_b^i.x_i^{(t)} &= r P_b^i.x_i^{(t-1)} + (1 - r) P_a^i.x_i^{(t-1)} \label{eq:cross_xb} 
\end{align}
\noindent where $r$ is a random number from the range $[0, 1]$. If $| P_a.v_i^{(t-1)} + P_b.v_i^{(t-1)} | \neq 0$, then their velocities are also updated using the following crossover operations (Lines~\ref{line:72}-\ref{line:73}):

\begin{align}
P_a^i.v_i^{(t)} &= \Bigg ( \frac{P_a^i.v_i^{(t-1)} + P_b^i.v_i^{(t-1)}}{|{P_a^i.v_i^{(t-1)} + P_b^i.v_i^{(t-1)} |}} \Bigg ) | P_a^i.v_i^{(t-1)} |
\label{eq:cross_va} \\
P_b^i.v_i^{(t)} &= \Bigg ( \frac{P_a^i.v_i^{(t-1)} + P_b^i.v_i^{(t-1)}}{|{P_a^i.v_i^{(t-1)} + P_b^i.v_i^{(t-1)} |}} \Bigg ) | P_b^i.v_i^{(t-1)} | 
\label{eq:cross_vb}
\end{align}
\noindent Otherwise, the velocities are updated using the regular update operations described in Equations~\ref{eq:4} and~\ref{eq:3}.

\subsection{Example Partial Trace}
\noindent
We now provide a partial trace of our PCD algorithm on the example C-DCOP of Figure~\ref{fdcopex}. Assume that the number of particles $K=4$. In the \textbf{initialization phase}, the agents cooperatively build the BFS pseudo-tree shown in Figure~\ref{bfsptree}, after which each agent $a_i$ is aware of its set of neighboring agents $N_i$, its set of children agents $CH_i$, and its parent agent $PR_i$: 

\begin{align}
    N_1 &= \{ a_2, a_3, a_4\}; CH_1 = \{ a_2, a_3, a_4\}; PR_1 = \emptyset\\
    N_2 &= \{a_1\}; CH_2 = \emptyset; PR_2 = a_1\\
    N_3 &= \{a_1, a_4\}; CH_3 = \emptyset; PR_3 = a_1\\
    N_4 &= \{a_1, a_3\}; CH_4 = \emptyset; PR_4 = a_1
\end{align}
Each agent $a_i$ then creates a set of particles $P^i = \{P^i_1, P^i_2, P^i_3, P^i_4\}$ and initializes their position and velocity attributes $P^i_k.x_i$ and $P^i_k.v_i$ for all particles $P^i_k \in P^i$. Assume that they are initialized using the assignments below:

\begin{align}
P_1.X &= \{x_1 = -1.0, x_2 = 1.2, x_3 = -2.0, x_4 = 2.0\}\\ 
P_2.X &= \{x_1 = -2.0, x_2 = 2.0, x_3 = -1.0, x_4 = 1.0\}\\ 
P_3.X &= \{x_1 = 0.0, x_2 = 1.0, x_3 = 2.0, x_4 = -2.0\}\\ 
P_4.X &= \{x_1 = 1.1, x_2 = -1.0, x_3 = 1.5, x_4 = 0.5\} \\
P_1.V &= P_2.V = P_3.V = P_4.V = \{v_1 = 0.0, v_2 = 0.0, v_3 = 0.0, v_4 = 0.0\}
\end{align}
Then, each agent $a_i$ sends its position assignments $P^i.x_i$ in a VALUE message to each of its neighboring agents in $N_i$:
\begin{itemize}
\item Agent $a_1$ sends a VALUE($P^1.x_1$) message to each of its neighboring agents $a_2$, $a_3$, and $a_4$, where $P^1.x_1 = \{P^1_1.x_1, P^1_2.x_1, P^1_3.x_1, P^1_4.x_1\} = \{-1.0, -2.0, 0.0, 1.1\}$.
\item Agent $a_2$ sends a VALUE($P^2.x_2$) message to its neighboring agent $a_1$, where $P^2.x_2 = \{P^2_1.x_2, P^2_2.x_2, P^2_3.x_2, P^2_4.x_2\} = \{1.2, 2.0, 1.0, -1.0\}$.
\item Agent $a_3$ sends a VALUE($P^3.x_3$) message to each of its neighboring agents $a_1$ and $a_4$, where $P^3.x_3 = \{P^3_1.x_3, P^3_2.x_3, P^3_3.x_3, P^3_4.x_3\} = \{-2.0, -1.0, 2.0, 1.5\}$.
\item Agent $a_4$ sends a VALUE($P^4.x_4$) message to each of its neighboring agents $a_1$ and $a_3$, where $P^4.x_4 = \{P^4_1.x_4, P^4_2.x_4, P^4_3.x_4, P^4_4.x_4\} = \{2.0, 1.0, -2.0, 0.5\}$.
\end{itemize}

In the \textbf{evaluation phase}, each agent $a_i$ waits for the VALUE messages from its neighboring agents. Upon receiving the VALUE messages from \emph{all} of its neighboring agents, it calculates the local fitness value $P^i_k.local\_fitness$ for each particle $P^i_k \in P^i$. For example, after receiving VALUE($P^1.x_1$) and VALUE($P^3.x_3$) from agents $a_1$ and $a_3$, respectively, agent $a_4$ calculates $P^4_1.local\_fitness$ for particle $P^4_1$ as follows (see Figure~\ref{fdcopex} for the set of cost functions of our example C-DCOP):

\begin{align}
P^4_1.local\_fitness &= f_{14}\left(P^1_1.x_1, P^4_1.x_4 \right) + f_{34}\left( P^3_1.x_3, P^4_1.x_4 \right) \\ 
&= 2 \left(P^1_1.x_1 \right) ^{ 2 } - 2 \left( P^4_1.x_4 \right) ^{ 2 } + \left( P^3_1.x_3 \right) ^{ 2 } + 3 \left( P^4_1.x_4 \right) ^{ 2 } \\ 
&= 2 \left( -1 \right) ^{ 2 } - 2 \left( 2 \right) ^{ 2 } + \left( -2 \right) ^{ 2 } + 3 \left( 2 \right) ^{ 2 }\\ &= 10 \;
\end{align}
Table~\ref{tab:worked_fitness} tabulates the values of $P^i_k.local\_fitness$ for each particle $P^i_k$ of each agent $a_i$. 

\begin{table}[]
\centering
\caption{Local Fitness Scores}
\begin{tabular}{@{}ccccc@{}}
\toprule
               & Agent $a_1$ & Agent $a_2$ & Agent $a_3$ & Agent $a_4$ \\ \midrule
Particle $P_1$ & -1.44       & -0.44       & 21.00          & 10.00          \\ 
Particle $P_2$ & 14.00          & 0.00         & 12.00          & 10.00          \\ 
Particle $P_3$ & -9.00          & -1.00          & 16.00          & 8.00           \\
Particle $P_4$ & 6.64        & 0.21        & 7.51        & 4.92        \\ \bottomrule
\end{tabular}
\label{tab:worked_fitness}
\end{table}

\begin{table}[]
\centering
\caption{Fitness Scores}
\begin{tabular}{@{}ccccc@{}}
\toprule
               & Agent $a_1$ & Agent $a_2$ & Agent $a_3$ & Agent $a_4$ \\ \midrule
Particle $P_1$ & 14.56       & -0.44       & 21.00       & 10.00       \\ 
Particle $P_2$ & 18.00       & 0.00        & 12.00       & 10.00       \\ 
Particle $P_3$ & 7.00        & -1.00       & 16.00       & 8.00        \\
Particle $P_4$ & 9.60        & 0.21        & 7.51        & 4.92        \\ \bottomrule
\end{tabular}
\label{tab:worked_fitness2}
\end{table}

After computing the local fitness values, since agent $a_4$ does not have any child agent, it assigns its local fitness value $P^4_k.local\_fitness$ of each particle $P^4_k \in P^4$ to that particle's regular fitness value $P^4_k.fitness$ and sends that information to its parent agent $a_1$ in a COST message. Similarly, agents $a_2$ and $a_3$ also do the same as they too do not have any child agent. Table~\ref{tab:worked_fitness2} tabulates the values of $P^i_k.fitness$ for each particle $P^i_k$ of each agent $a_i$, and the COST messages sent by the agents are below:
\begin{itemize}
\item Agent $a_2$ sends a COST($P^2.fitness$) message to its parent agent $a_1$, where $P^2.fitness = \{P^2_1.fitness, P^2_2.fitness, P^2_3.fitness, P^2_4.fitness\} = \{-0.44, 0.00, -1.00, 0.21\}$.
\item Agent $a_3$ sends a COST($P^3.fitness$) message to its parent agent $a_1$, where $P^3.fitness = \{P^3_1.fitness, P^3_2.fitness, P^3_3.fitness, P^3_4.fitness\} = \{21.00, 12.00, 16.00, 7.51\}$.
\item Agent $a_4$ sends a COST($P^4.fitness$) message to its parent agent $a_1$, where $P^4.fitness = \{P^4_1.fitness, P^4_2.fitness, P^4_3.fitness, P^4_4.fitness\} = \{10.00, 10.00, 8.00, 4.92\}$.
\end{itemize}
As the root agent $a_1$ has children agents, it waits for the COST messages from its children agents. Upon receiving the COST messages from \emph{all} of its children agents, it calculates the fitness value $P^1_k.fitness$ for each particle $P^1_k \in P^1$. For example, it calculates $P^1_1.fitness$ for particle $P^1_1$ as follows:
\begin{align}
P^1_1.fitness &= P^1_1.local\_fitness + P^2_1.fitness + P^3_1.fitness + P^4_1.fitness \\
&= -1.44 + (-0.44) + 21.00 + 10.00 \\
&= 29.12
\end{align}
As the cost from each cost function in the C-DCOP is doubly counted, the root agent divides its fitness value of each of its particles by two. For example, it updates $P^1_1.fitness$ for particle $P^1_1$ as follows:
\begin{align}
P^1_1.fitness &= \frac{1}{2} P^1_1.fitness = \frac{1}{2}29.12 = 14.56
\end{align}

In the \textbf{update phase}, since this is the first iteration, each particle $P^1_k$ of the root agent $a_1$ has a better local position. Thus, the best position $P^1_k.p_{best}$ of each particle $P^1_k$ is updated to the particle's current position $P^1_k.x_1$ and all the particles are added into the set $PB$. Similarly, a better global position is found. Thus, the best global position $P^1.g_{best}$ is updated to the position $P_3^1.x_1$ of the best particle $P_3^1$ and that particle is assigned to the variable $P^*$. The agent then sends both $PB$ and $P^*$ in a BEST message to its children agents $a_2$, $a_3$, and $a_4$: 
\begin{itemize}
\item Agent $a_1$ sends a BEST($PB$, $P^*$) message to its children agents $a_2$, $a_3$, and $a_4$, where $PB = \{P^1_1, P^1_2, P^1_3, P^1_4\}$ and $P^* = P^1_3$.
\end{itemize}

All non-root agents $a_2$, $a_3$, and $a_4$ wait for the BEST messages from their parent agent $a_1$. Upon receiving the BEST message, for each particle $P_k \in PB$ in the BEST message, each agent assigns the position $P_k.x_i$ as the best local position $P_k^i.p_{best}$ of the corresponding particle $P_k^i$. Since $PB$ contains all four particles, the best local positions of all four particles are updated. Similarly, each agent $a_i$ also updates the best global position $P^i.g_{best}$ to the position $P^*.x_i = P_3^i.x_i$ of the best particle $P^*$ in the BEST message. 
Table~\ref{tab:best_upd} tabulates the values of the best local positions $P^i_k.p_{best}$ for each particle $P^i_k$ and the best global position $P^i.g_{best}$ of each agent $a_i$.

\begin{table}[]
    \centering
    \caption{$P^i_k.p_{best}$ and $P^i.g_{best}$ of Agent $a_i$ and Particle $P^i_k$ after BEST Update \label{tab:best_upd}}
    \begin{subtable}[c]{\textwidth}
        \centering
        \caption{$P_k^i.p_{best}$ for each Agent $a_i$ \label{subtab:pbest}}
        \begin{tabular}{@{}ccccc@{}}
\toprule
                                & Agent $a_1$           & Agent $a_2$           & Agent $a_3$           & Agent $a_4$           \\ \midrule
Particle $P_1$ & -1.00 & 1.20 & -2.00 & 2.00 \\

Particle $P_2$ & -2.00 & 2.00 & -1.00 & 1.00 \\
                                
Particle $P_3$ & 0.00 & 1.00 & 2.00 & -2.00 \\
                               
Particle $P_4$                  & 1.10 & -1.00 & 1.50 & 0.50
                                \\ \bottomrule
\end{tabular}
        
        \vspace{5mm}           
    \end{subtable}
\quad%
    \begin{subtable}[c]{\textwidth}
        \centering
        \caption{$P^i.g_{best}$ for each Agent $a_i$ \label{subtab:gbest}}
        \begin{tabular}{@{}ccccc@{}}
\toprule
                                & Agent $a_1$           & Agent $a_2$           & Agent $a_3$           & Agent $a_4$           \\ \midrule

Particle $P^*$  & 0.00 & 1.00 & 2.00 & -2.00 \\
                                \bottomrule
\end{tabular}
        \vspace{3mm}           
    \end{subtable}
\end{table}

Each agent $a_i$ then increments its cycle counter $t = 1$ before starting the process of updating the positions $P^i_k.x_i$ and velocities $P^i_k.v_i$ of its particles $P^i_k$. To do so, each agent first calculates $s_c$, $f_c$, and $\rho$ using Equations~\ref{eq:7},~\ref{eq:8}, and~\ref{eq:6}, respectively. In the following, assume that we set $max_{s_c} = 15$, $max_{f_c} = 5$, $w = 0.72$, $c_1 = 1.49$, $c_2 = 1.49$.\footnote{We discuss the choice of parameter values in detail in Section~\ref{sec:param_choice}} Since a new $P^*$ is found in this cycle, the agent sets $s_c^{(1)} = 1$ (see Equation~\ref{eq:7}) and $f_c^{(1)} = 0$ (see Equation~\ref{eq:8}). And since $0 = s_c^{(0)} \not > max_{s_c} = 15$ and $0 = f_c^{(0)} \not > max_{f_c} = 5$, $\rho^{(1)} = \rho^{(0)} = 1$ (see Equation~\ref{eq:6}), each agent $a_i$ updates the positions $P^i_k.x_i$ and velocities $P^i_k.v_i$ of its particles $P^i_k$ using Equations~\ref{eq:4} to~\ref{eq:5}. For example, agent $a_1$ updates the position and velocity of its best global particle $P^1_3$ as follows:

\begin{align} 
P^1_3.v_1^{(t)} & = -P^1_3.x_1^{(0)} + P^1.g_{best}^{(0)} + w P^1_3.v_1^{(0)} + \rho^{(1)}(1-2r_2) \\
& = 0 + 0 + 0.72 \cdot 0 + 1 (1-2\cdot0.4)\\
& = 0.20\\
P^1_3.x_1^{(1)} &= P^1_3.x_1^{(0)} + P^1_3.v_1^{(1)} = 0 + 0.2 = 0.20
\end{align}
and the position and velocity of a non-best global particle $P_1^1$ as follows:

\begin{align} 
P_1^1.v_1^{(1)} &= w P_1^1.v_1^{(0)} +  r_1 c_1  (P_1^1.p_{best}^{(0)} - P_1^1.x_1^{(0)}) 
 + r_2  c_2  (P^1.g_{best}^{(0)} - P_1^1.x_1^{(0)}) \\
& = 0.72 \cdot 0 + 0.7 \cdot 1.49 (-1-(-1)) + 0.4 \cdot 1.49 (0-(-1)) \\
& = 0.60\\
P^1_1.x_1^{(1)} &= P^1_1.x_1^{(0)} + P^1_1.v_1^{(1)} = -1 + 0.60 = -0.40
\end{align}
Table~\ref{tab:variable_upd} tabulates the updated position $P^i_k.x_i^{(1)}$ and velocity $P^i_k.v_i^{(1)}$ for each particle $P^i_k$ of each agent $a_i$.

\begin{table}[t]
\centering
\caption[The Variable Values after Variable Update ]{%
  Updated Particle Position and Velocity}
\begin{tabular}{@{}lllll@{}}
\toprule
                                & Agent $a_1$  & Agent $a_2$  & Agent $a_3$ & Agent $a_4$  \\ \midrule
\multirow{2}{*}{Particle $P_1$} & $v_1=$ 0.60   & $v_2=$ 1.19  & $v_3=$ 0.20  & $v_4=$ -0.66 \\
                                & $x_1=$ -0.40  & $x_2=$ -0.81 & $x_3=$ 0.20  & $x_4=$ 0.44  \\ \midrule
\multirow{2}{*}{Particle $P_2$} & $v_1=$ -0.12 & $v_2=$ -0.60  & $v_3=$ 0.20  & $v_4=$ 1.19  \\
                                & $x_1=$ 1.08  & $x_2=$ 1.40   & $x_3=$ 1.20  & $x_4=$ 0.19  \\ \midrule
\multirow{2}{*}{Particle $P_3$} & $v_1=$ 2.38  & $v_2=$ 1.79  & $v_3=$ 0.20  & $v_4=$ 0.30   \\
                                & $x_1=$ 0.38  & $x_2=$ 0.79  & $x_3=$ 2.20  & $x_4=$ 1.80   \\ \midrule
\multirow{2}{*}{Particle $P_4$} & $v_1=$ -2.38 & $v_2=$ -1.79 & $v_3=$ 0.20  & $v_4=$ -1.49 \\
                                & $x_1=$ -0.38 & $x_2=$ -0.79 & $x_3=$ -1.80 & $x_4=$ -0.99 \\ \bottomrule
\end{tabular}
\label{tab:variable_upd}
\end{table}

\begin{table}[]
\centering
\caption[Crossover Probabilities ]{%
  Crossover Probabilities }
\begin{tabular}{@{}ccccc@{}}
\toprule
              & Agent $a_1$ & Agent $a_2$ & Agent $a_3$ & Agent $a_4$ \\ \midrule
Particle $P_1$ & 0.046       & 0.267       & 0.372       & 0.304       \\ 
Particle $P_2$ & 0.450       & 0.000       & 0.212       & 0.304       \\ 
Particle $P_3$ & 0.290       & 0.606       & 0.283       & 0.243       \\ 
Particle $P_4$ & 0.214       & 0.127       & 0.133       & 0.149       \\ \bottomrule
\end{tabular}
\label{tab:worked_wp}
\end{table}

We now describe an example of the crossover operation in PCD\_CrossOver variant. Using the local fitness values calculated in Table~\ref{tab:worked_fitness}, each agent $a_i$ calculates its crossover probability $P_k^i.b_p$ for each of its particles $P_k^i$ using Equation~\ref{eq:bp}. For example, agent $a_1$ calculates $P^1_1.b_p$ for particle $P^1_1$ as follows:

\begin{align}
    P^1_1.b_p = \frac{| -1.44 |}{| -1.44 | + | 14 | + | -9 | + | 6.64 |} = 0.046
    \label{eq:worked_wp}
\end{align}
Table~\ref{tab:worked_wp} tabulates the crossover probabilities $P_k^i.b_p$ for each particle $P_k^i$ of each agent $a_i$.

Using these probabilities, each agent $a_i$ selects two random particles $P^i_a$ and $P^i_b$ and update their positions and velocities using Equations~\ref{eq:cross_xa} to~\ref{eq:cross_vb}. For example, agent $a_1$ selects particles $P^1_2$ and $P^1_4$ and updates their positions as follows:

\begin{align}
P_2^1.x_1^{(1)} &= r P_2^1.x_1^{(0)} + (1 - r) P_4^1.x_1^{(0)}\\
& = 0.3 \cdot -2 + (1-0.3) 1.1 \\
& = 0.17\\
P_4^1.x_1^{(1)} &= r P_4^1.x_1^{(0)} + (1 -r) P_2^1.x_1^{(0)} \\
 & = 0.3 \cdot 1.1 + (1-0.3) (-2) \\
 & = -1.07
\end{align}
where $r = 0.3$. Their velocities are not updated because $| P_2^1.v_1^{(0)} + P_4^1.v_i^{(0)} | = | 0 + 0 | = 0$.

\section{Theoretical Analysis}\label{section:theory}
\noindent
In this section, we first prove PCD is an anytime algorithm, that is, the quality of the best solution improves and never degrades over time. We then discuss the complexity of PCD in terms of its communication and memory requirements. In this section, we use the term \emph{iteration} to refer to communication steps, which is the time needed for messages sent by an agent to be received by its neighboring agents.

\begin{lemma}
At iteration $t+h$, the root agent $a_{root}$ is aware of the $P^{root}.p_{best}$ and $P^{root}.g_{best}$ up to iteration $t$, where $h$ is the height of the pseudo-tree.
\end{lemma}

\begin{sproof}
To prove the lemma, we show that, at iteration $t+h$, the root agent has enough information to calculate $P^{root}.p_{best}$ and $P^{root}.g_{best}$ up to iteration $t$. That is, the root agent can calculate the fitness of each particle. However, the root agent requires the COST messages from all the agents in $CH_{root}$ in order to calculate the fitness of each particle using Equation \ref{eq:2}.  To get these messages, the root agent has to wait for at most $h$ iterations since the height of the pseudo-tree is $h$. Consequently, at iteration $t+h$, the root  agent is capable of calculating the fitness of each particle up to iteration $t$. 
\end{sproof}

\begin{lemma}
At iteration $t+2h$, each agent $a_i$ is aware of $P^i.p_{best}$ and $P^i.g_{best}$ up to iteration $t$, where $h$ is the height of the pseudo-tree.
\end{lemma}

\begin{sproof}
It will take at most $h$ iterations for the BEST message that contains $(PB, P^*)$ from the root agent to reach all other agents since $h$ is the height of the pseudo-tree. Therefore, combining this observation and Lemma 1, it will take at most $t+h+h=t+2h$ iterations for each agent to be aware of $P.p_{best}$ and $P.g_{best}$ up to iteration $t$. 
\end{sproof}

\begin{theorem}
PCD is an anytime algorithm.
\end{theorem}

\begin{sproof}
From Lemma 2, at iteration $t+2h$ and $t+2h+\delta$, where $\delta \geq 0$, each agent is aware of $P^i.p_{best}$ and $P^i.g_{best}$ up to iterations $t$ and $t+\delta$, respectively. Since $P^i.p_{best}$ and $P^i.g_{best}$ only gets updated if a better solution is found, $P^i.p_{best}.fitness$ and $P^i.g_{best}.fitness$ at iteration $t+2h+\delta$ is no larger than $P^i.p_{best}.fitness$ and $P^i.g_{best}.fitness$ at iteration $t+2h$, respectively. In other words, the cost of the solution is monotonically non-increasing over time. Hence, PCD is an anytime algorithm. 
\end{sproof}

\begin{theorem}
For a binary constraint graph $G = (|A|, E)$, the number of messages of PCD in each cycle is $O(|E| + |A|)$.
\end{theorem}

\begin{sproof}
In each cycle: 
\begin{itemize}
    \item The \texttt{INITIALIZATION} procedure requires $2|E|$ messages, where $|E|$ is the number of edges in the constraint graph, since each agent sends a VALUE message to each of its neighbors. 
    \item The \texttt{EVALUATION} procedure requires $|A|-1$ messages since each agent, except for the root agent, sends a COST message to its parent agent.
    \item The \texttt{BEST\_UPDATE} procedure requires $|A|-1$ messages since each agent, except for the root agent, receives a BEST message from its parent agent.
    \item The \texttt{VARIABLE\_UPDATE} procedure requires $2|E|$ messages since each agent sends a VALUE message to each of its neighbors. 
\end{itemize}
Since each cycle of PCD is composed of either the \texttt{INITIALIZATION} or \texttt{VARIABLE\_UPDATE} procedures -- \texttt{INITIALIZATION} procedure in the first cycle and \texttt{VARIABLE\_UPDATE} in subsequent cycles -- as well as the \texttt{EVALUATION} and \texttt{BEST\_UPDATE} procedures, the number of messages per cycle is thus $2|E| + |A| - 1 + |A| - 1 = O(|E| + |A|)$.
\end{sproof}

\begin{theorem}
The total message size complexity per agent of PCD in each cycle is $O(K|A|)$, where $K$ is the total number of particles.
\end{theorem}

\begin{sproof}
Each agent in PCD sends three types of messages: VALUE, COST, and BEST messages. Each of these messages contains a constant amount of information for each of the $K$ particles. Hence, the size of each message is $O(K)$. At each cycle, each agent $a_i$ sends at most $|N_i|$ VALUE messages, one COST message, and $|CH_i|$ BEST messages. Therefore, the total number of messages an agent sends is at most $|N_i|+1+|CH_i|$. In the worst case, $|N_i| \approx |A|$ and $|CH_i| \approx |A|$. Consequently, the total number of message sent by an agent is $O(|A|+1+|A|) = O(|A|)$.  Therefore, in each cycle, the total message size per agent is $O(K|A|)$.
\end{sproof}

\section{Experimental Results}\label{section:exp}
\noindent
We now provide our results from empirical evaluations of PCD. First, we study the effect of different parameter choices to fine tune the parameters. This is because the performance of PCD depends on the choice of its various parameter values and whether crossover operations are used to improve the algorithm further. Then, we compare the fine-tuned version of PCD against the state-of-the-art C-DCOP algorithms\footnote{Although \citeauthor{hoang2020new}~\cite{hoang2020new} proposed three non-exact C-DCOP algorithms, we only compare with AC-DPOP and C-DSA here because they are reported to provide the best solutions among the approximate algorithms proposed in their paper.} -- namely HCMS~\cite{voice2010hybrid}, AC-DPOP~\cite{hoang2020new}, and C-DSA~\cite{hoang2020new} -- on four benchmark problems -- random graphs, random trees, scale-free networks, and random sensor network problems. 

\subsection{Benchmark Problems}
\label{sec:exp_config}
\noindent
We evaluate PCD on four types of benchmark problems: \textit{Random Graphs}, \textit{Random Trees}, \textit{Scale-Free Networks}, and \textit{Random Sensor Network Problems}.

\smallskip \noindent \textbf{Random Graphs:} We use the Erd{\H{o}}s-R{\'e}nyi topology~\cite{erdHos1960evolution} to construct random graphs. We use two settings for random graphs -- sparse, where each pair of nodes in the graph has a probability of 0.2 to have an edge between them, and dense, where the edge probability is 0.6. We set the number of agents $|A|$, which corresponds to the number of nodes in the graph, to 50 and the domain $D_i$ of each agent $a_i$ to $[-50, 50]$. We use binary quadratic functions of the form $ax^2 + bxy + cy^2$ as cost functions, where the coefficients $a$, $b$, and $c$ are all randomly chosen between $[-5, 5]$.

\smallskip \noindent \textbf{Random Trees:} We follow \citeauthor{hoang2020new}~\cite{hoang2020new} and include  random trees as one of the benchmark problems since the memory requirement of AC-DPOP is smaller on trees; it is exponential in the tree width of the graph~\cite{hoang2020new}.  The experimental configurations are similar to the random graph setting, except that we ensure that no cycles are formed.

\smallskip \noindent \textbf{Scale-Free Networks:} In scale-free networks, the problem settings are similar to Random
Graphs, except we use the Barabsi-Albert (BA) network topology model \cite{barabasi1999emergence} to generate our constraint networks. To construct the network, initially, we randomly connect a set of agents. We then connect a new agent with a set of $m$ randomly selected existing agents with a probability proportional to the current link numbers. For our problem, we set number of agents $|A| = 100$ and $m = 3$ and the domain $D_i$ of each agent $a_i$ to $[-20, 20]$. Other experimental configurations are same as the random graph setting. The purpose of adding this experiment is to observe the impact of different network topologies (random graph, scale-free and random tree) and larger number of agents on the performance of PCD and PCD\_CrossOver.

\smallskip \noindent \textbf{Random Sensor Network Problems:} The fourth benchmark problem is motivated by an example sensor network problem, where the sensors are trying to maximize the signal strength of their radio communication~\cite{nguyen2012stochastic, nguyen2019distributed}. This problem is motivated by real-world applications where a \emph{meshed communication network} needs to be established by first responders of an emergency rescue operation (e.g., in a city ravaged by an earthquake or a hurricane) or by soldiers in the battlefield. In such a problem, the sensors are arranged in an $8 \times 8$ rectangular grid and can make small movements within its cell in the grid. Therefore, the number of agents (i.e., sensors) $|A|$ is $64$. The domain $D_i$ of each agent $a_i$ is set to the cell size $[0,10]$.

The strength of the radio communication between two sensors $a_i$ and $a_j$ is inversely proportional to their squared distance $d(a_i,a_j)^2$ and is subject to interference $\lambda(a_i,a_j)$ from obstacles between the sensors. We thus define the utility $f(a_i, a_j)$ between two sensors $(a_i, a_j)$ as follows:

\begin{align}
    f(a_i, a_j) &= \frac{C}{d(a_i, a_j)^2 \cdot \lambda(a_i,a_j)}\\
    \lambda(a_i,a_j) &= (r_{x_i}-x_i)^2 + (r_{y_i}-y_i)^2 + (r_{x_j} - x_j) ^ 2 + (r_{y_j} - y_j)^2 + \eta_{ij}
\end{align}
where $r_{x_i}$ and $r_{y_i}$ are random numbers sampled from the domain of agent $a_i$, $\eta_{ij}$ is random noise chosen between the range $[1,10]$ for each pair of sensors $a_i$ and $a_j$, and $C$ is a constant value of 10,000.




In all of the settings described above, we evaluate all the algorithms on 25 independently-generated problem instances and 20 times on each problem instance. For fairness, we use the simulated runtime metric~\cite{sultanik2008dcopolis} to measure the runtime of the algorithms. The experiments are carried out on a computer with an Intel Core i5-6200U CPU with 2.3GHz processor and 8GB RAM.

\subsection{Fine-tuning Parameters}
\label{sec:param_choice}

\begin{figure}[t]
\centering
\begin{subfigure}{0.7\textwidth}
  \centering
  \includegraphics[width=1\linewidth]{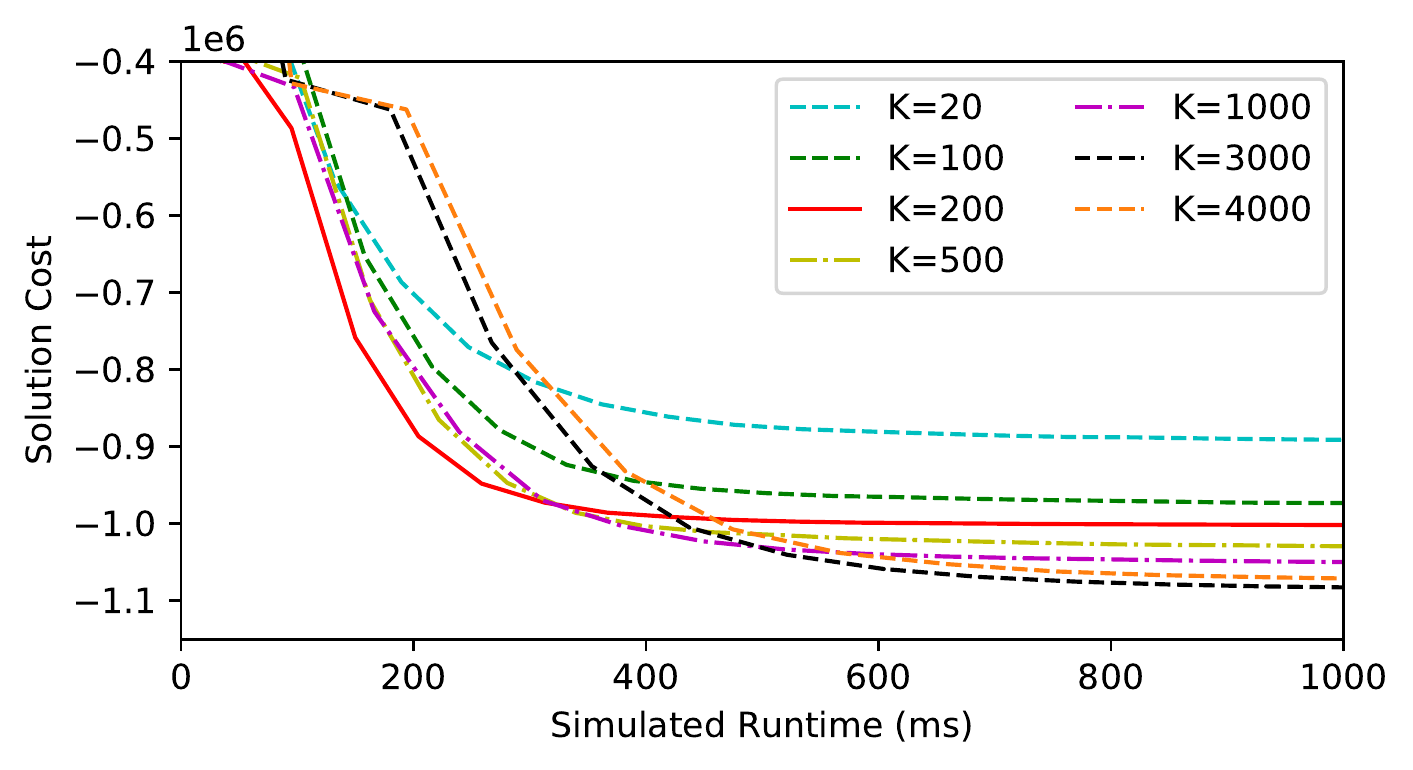}
  \caption{PCD}
  \label{fig:param_sparse_k1}
\end{subfigure}%

\begin{subfigure}{.7\textwidth}
  \centering
  \includegraphics[width=1\linewidth]{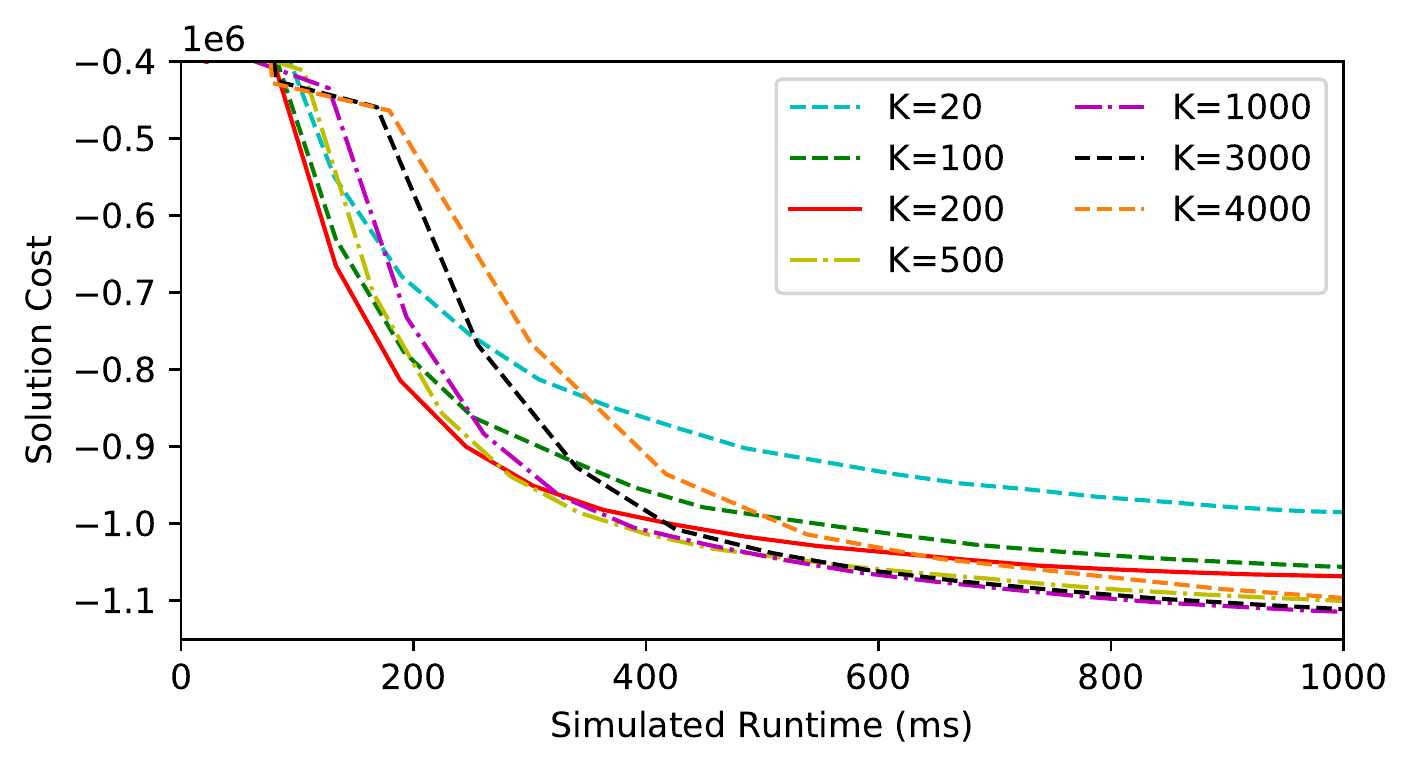}
  \caption{PCD\_CrossOver}
  \label{fig:param_sparse_k2}
\end{subfigure}
\caption{Solution quality of PCD and PCD\_CrossOver with different population sizes $K$ on sparse random graphs.}
\label{fig:param_sparsek}
\end{figure}

\begin{figure}[t]
\centering
\begin{subfigure}{0.75\textwidth}
  \centering
  \includegraphics[width=1\linewidth]{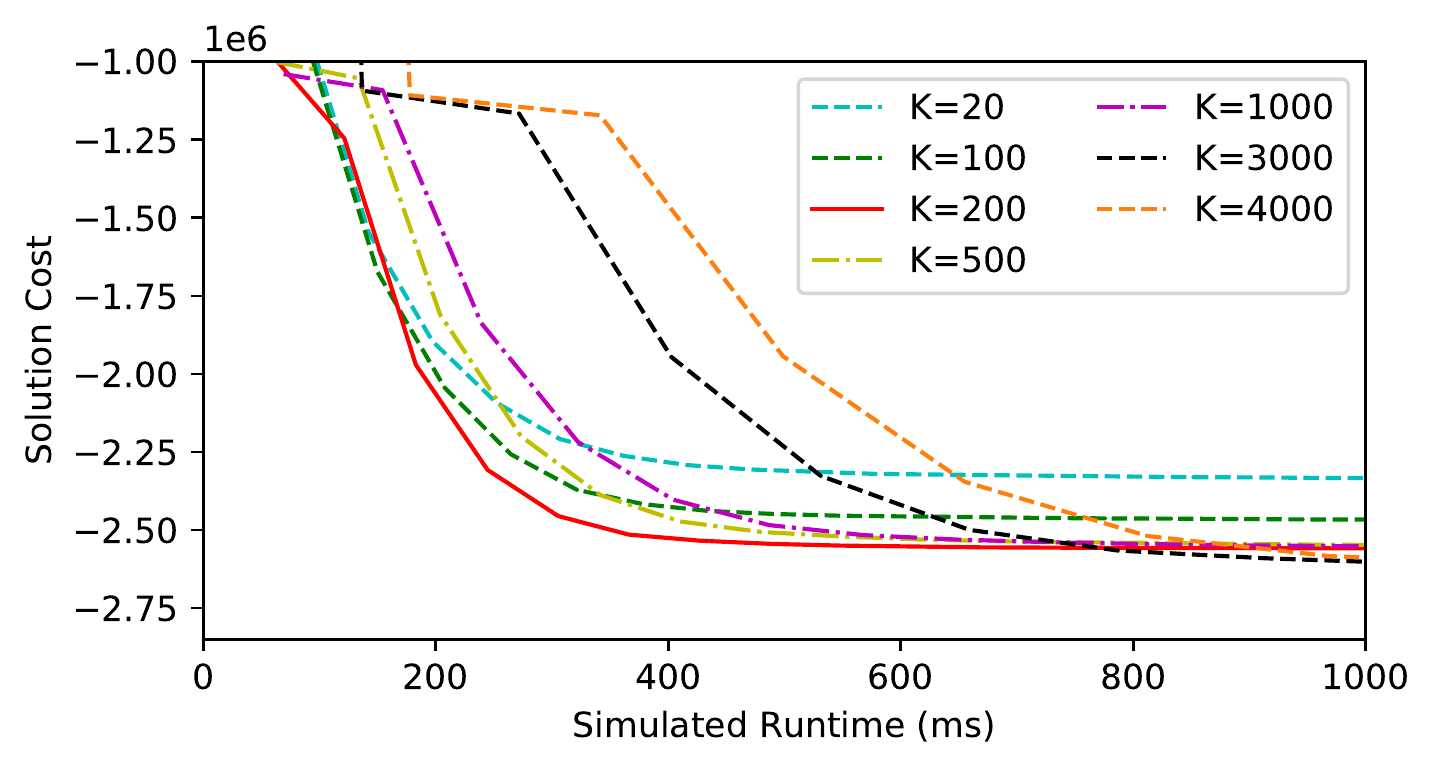}
  \caption{PCD}
  \label{fig:param_dense_k1}
\end{subfigure}%

\begin{subfigure}{.75\textwidth}
  \centering
  \includegraphics[width=1\linewidth]{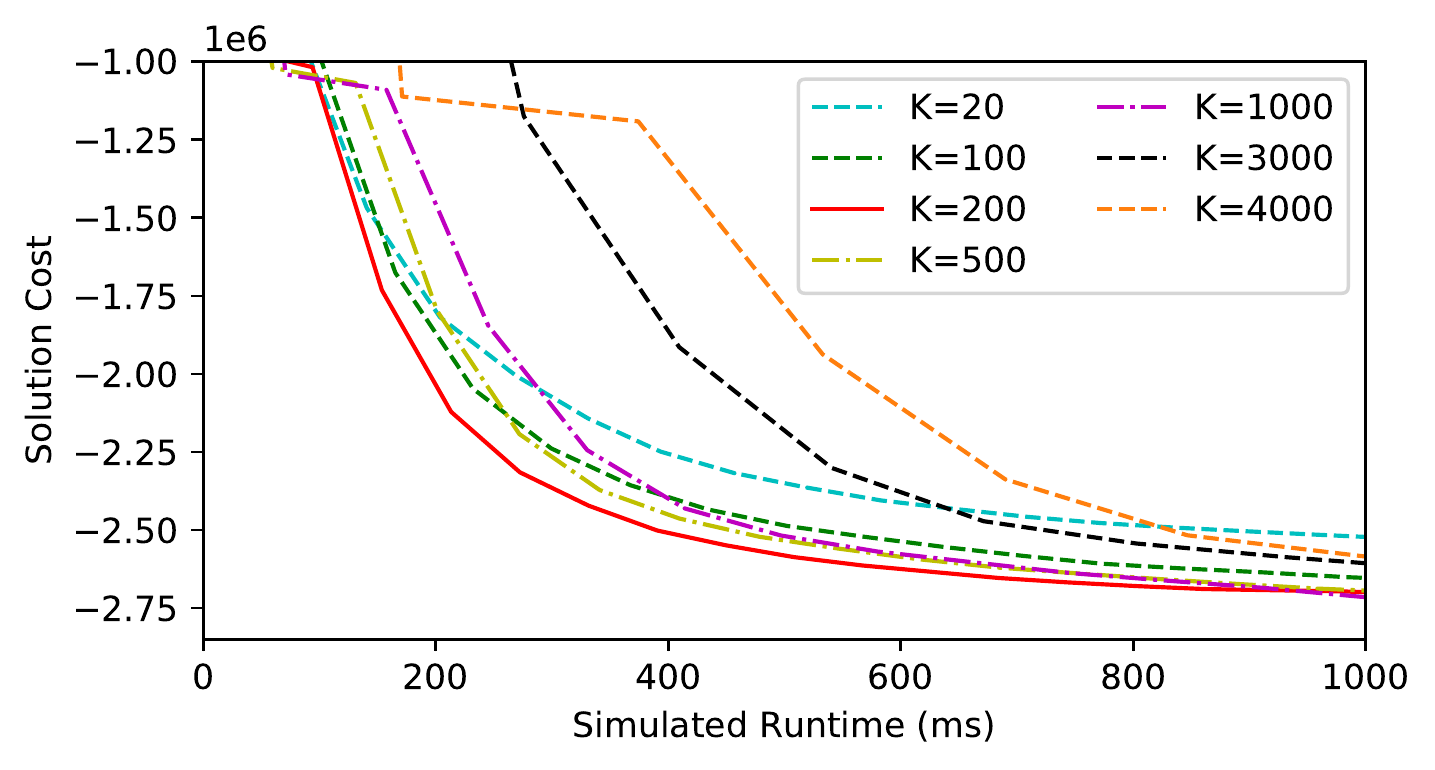}
  \caption{PCD\_CrossOver}
  \label{fig:param_dense_k2}
\end{subfigure}
\caption{Solution quality of PCD and PCD\_CrossOver with different population sizes $K$ on dense random graphs.}
\label{fig:param_densek}
\end{figure}

\noindent
PCD and its variant PCD\_CrossOver have several parameters including the number of particle $K$, inertia weight $w$, cognitive constant $c_1$, social constant $c_2$, and thresholds $max_{f_{c}}$ and $max_{s_{c}}$. In all our experiments, we follow the recommendations from the literature~\cite{van2002new} and set $max_{f_{c}} = 5$ and $max_{s_{c}} = 15$. We now discuss how we choose the other parameter values. 

To determine the value of the number of particles $K$, we conduct a preliminary experiment, where we compare the quality of solutions found by PCD and PCD\_CrossOver for different values of $K$. Figures~\ref{fig:param_sparsek} and~\ref{fig:param_densek} plot the results on sparse and dense random graphs, respectively. In general, the quality of solutions improves with increasing $K$ since a larger population size allows for more diversity in the swarm. However, this comes at the cost for longer runtimes before the algorithms converge. Given these empirical observations, we choose to set $K = 200$ in all our subsequent experiments as it allows the algorithms to converge to reasonably good solutions -- it is in fact very close to the best solutions with $K=4000$ on dense random graphs -- as well as converge to them relatively quickly compared to the other values of $K$.
\begin{figure}[t]
\centering
\begin{subfigure}{0.7\textwidth}
  \centering
  \includegraphics[width=1\linewidth]{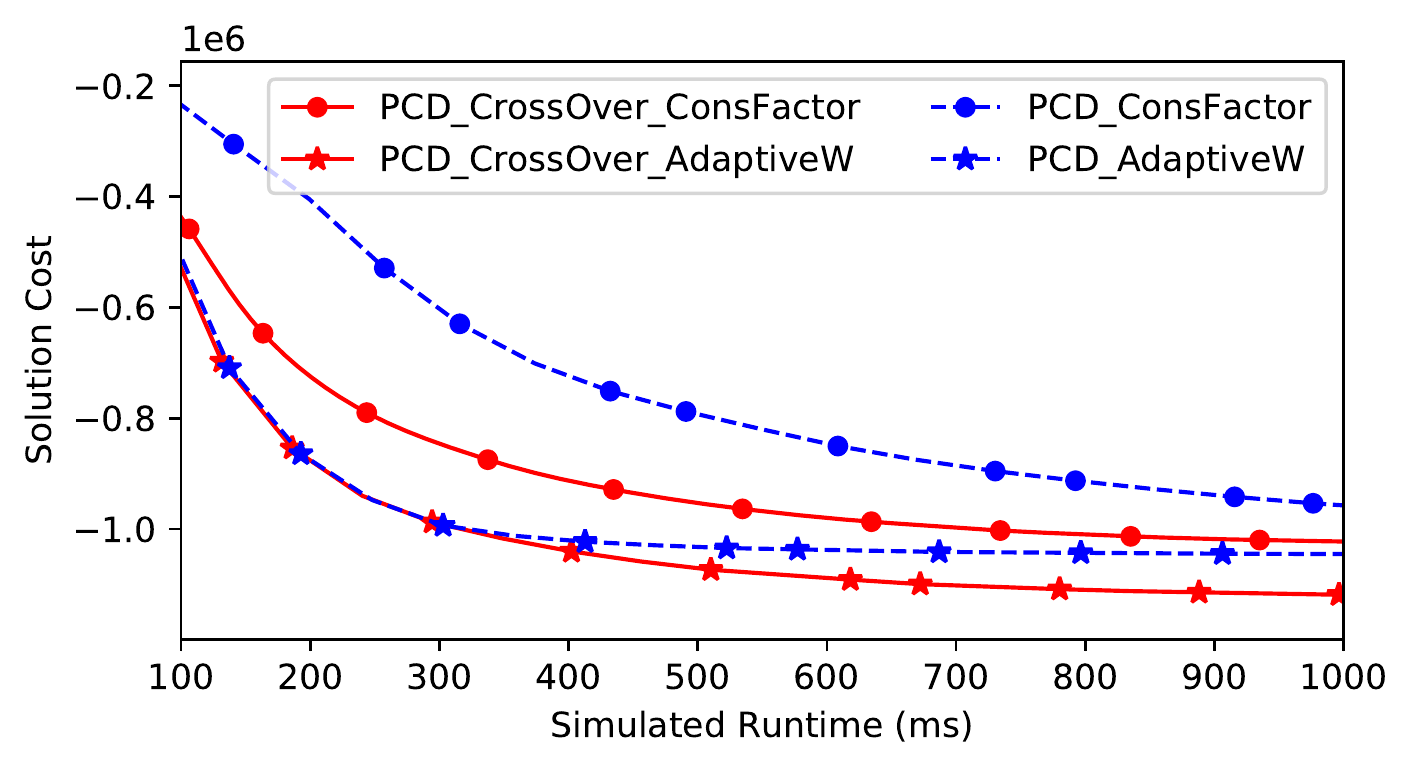}
  \caption{Sparse graph}
  \label{fig:param_sub1}
\end{subfigure}%

\begin{subfigure}{.7\textwidth}
  \centering
  \includegraphics[width=1\linewidth]{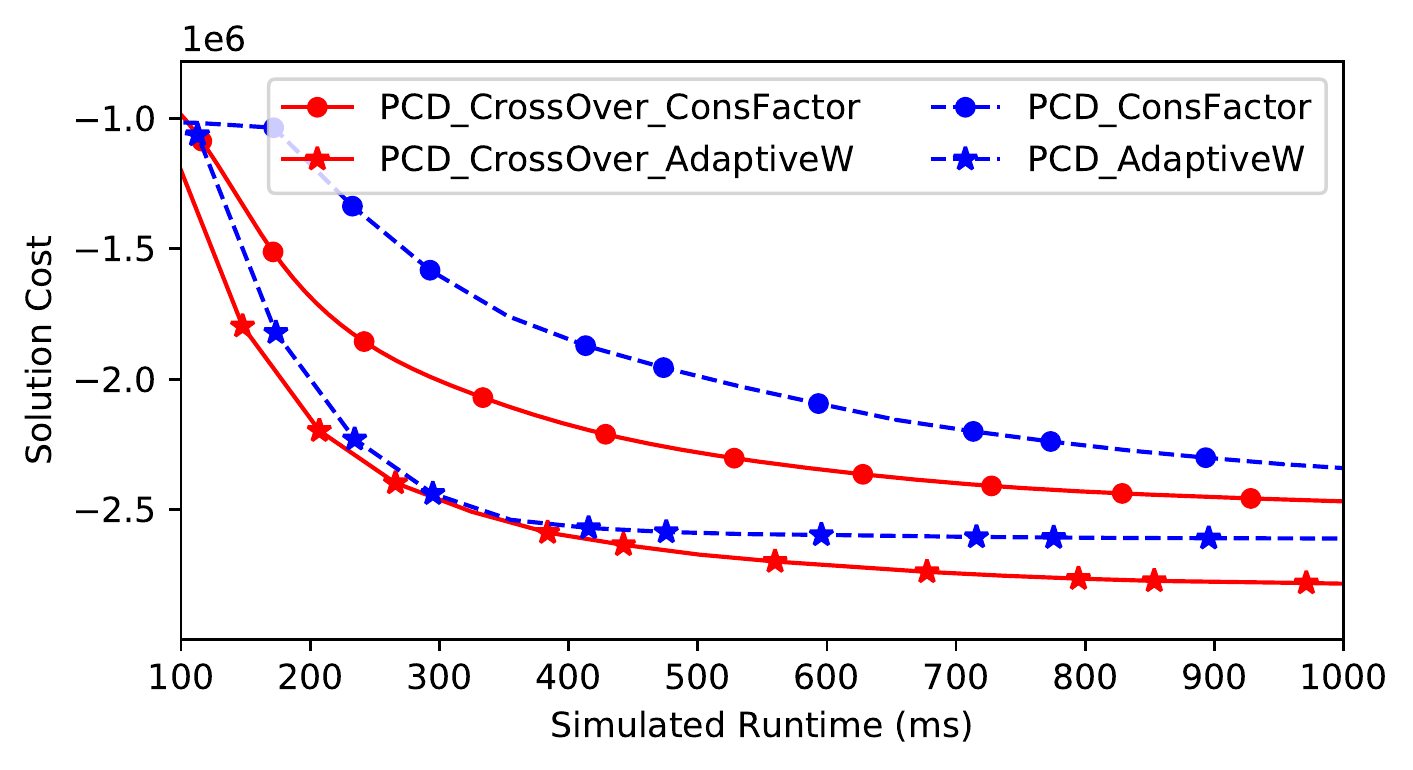}
  \caption{Dense graph}
  \label{fig:param_sub2}
\end{subfigure}
\caption{Solution quality of PCD and PCD\_CrossOver with AdaptiveW or Constriction Factor on random graphs.}
\label{fig:param}
\end{figure}
To determine the values of $w$, $c_1$, and $c_2$, we use two design choices: 
\begin{itemize}
\item \textbf{AdaptiveW:} Here, we linearly decrease $w$ using Equation \ref{eq:decW}:

\begin{align}
    w = \frac{(w_{max} - w_{min}) \cdot t}{t_{max}}
\label{eq:decW}
\end{align}
where $w_{max}$ and $w_{min}$ are the maximum and minimum values of $w$ and $t_{max}$ is the maximum number of cycles. To determine the values of $w_{max}$ and $w_{min}$, we experimented with different pairs of values typically used in experiments of centralized PSO \cite{shi1999empirical,shi1998modified,carlisle2000adapting}, and found that the best results are when $w_{max} = 1.4$ and $w_{min}=0.4$. For the $c_1$ and $c_2$ values, we set them to $c_1 = c_2 = 1.49$, which have been a popular choice in the centralized PSO model~\cite{eberhart2000comparing, van2010convergence}. 

\item \textbf{Constriction Factor:} Here, we follow the literature~\cite{clerc1999swarm}, where, instead of using Equation \ref{eq:3} to update the velocity of particles, we use Equation \ref{eq:consUpd} instead:

\begin{align}
\!\!\!\!\!\!\!\!\!\!\!\!P^i_k.v_i^{(t)} &= w(P^i_k.v_i^{(t-1)} + 
 r_1 c_1(P^i_k.p_{best}^{(t-1)} - P^i_k.x_i^{(t-1)})
 + r_2 c_2(P^i.g_{best}^{(t-1)} - P^i_k.x_i^{(t-1)}))
\label{eq:consUpd} \\
w &= \frac{2}{2-\phi-\sqrt{(\phi^2 - 4\phi)}}
\label{eq:consUpd2} \\
\phi &= c_1 + c_2 > 4
\label{eq:consUpd3}
\end{align}

To satisfy the constraint in Equation~\ref{eq:consUpd3}, we choose $c_1 = c_2 = 2.05$ to get $\phi = 4.1$ and $w = 0.7298$ because this set of parameter values has been proven to be a convergent parameter configuration for the centralized GCPSO model~\cite{van2010convergence}.
\end{itemize}

We then conduct another preliminary experiment, where we compare these two design choices for both PCD and PCD\_CrossOver on random graphs. Figure~\ref{fig:param} plot the results. It is clear that the parameters values tuned using AdaptiveW result in better solutions compared to when they are tuned using the constriction factor approach for both PCD and PCD\_CrossOver and in both sparse and dense random graphs. Therefore, for all subsequent experiments, we use the AdaptiveW approach to tune the $w$, $c_1$, and $c_2$ parameter values.

\subsection{Comparisons with the State of the Art}
\label{sec:exp_performance}

\begin{figure}[t]
\centering
\begin{subfigure}{.75\textwidth}
  \centering
  \includegraphics[width=1\linewidth]{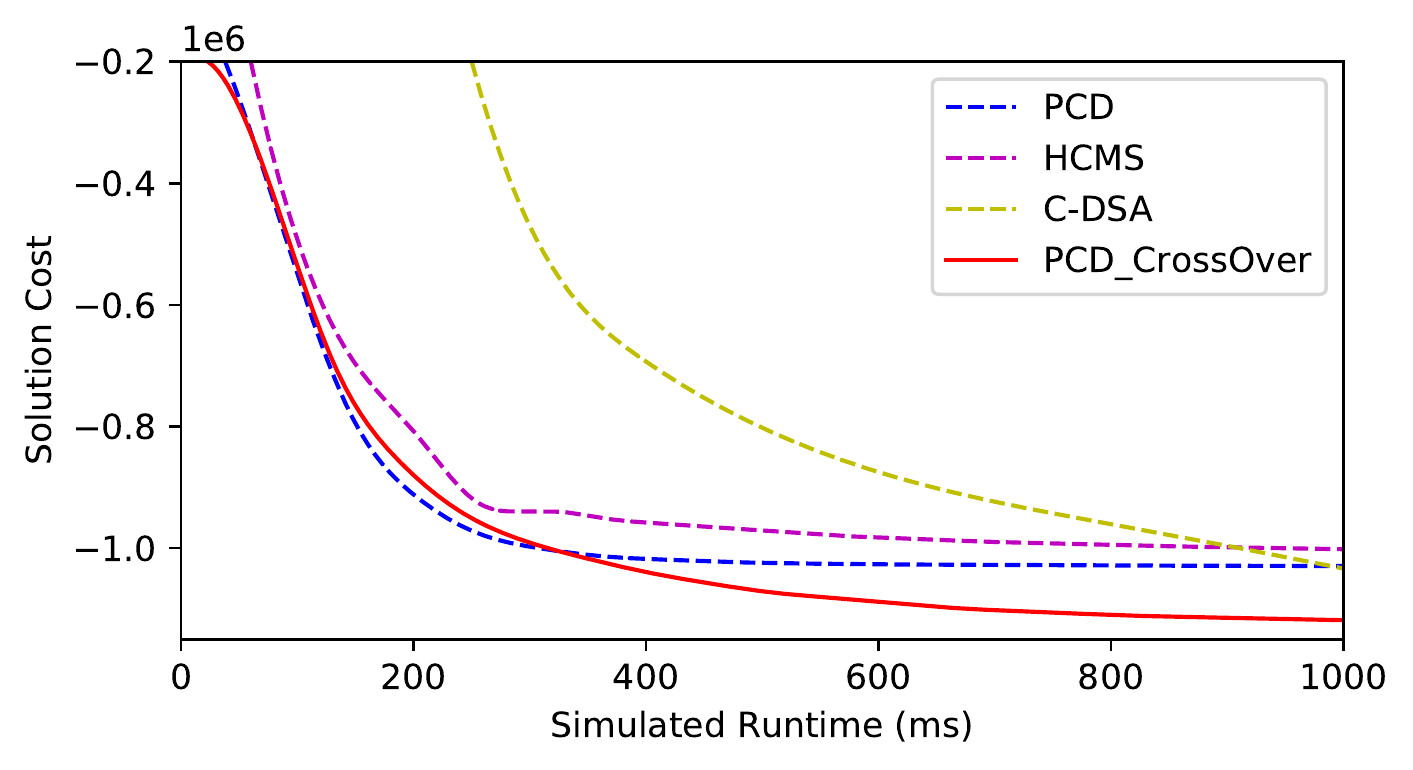}
  \caption{Sparse graph}
  \label{fig:sparse}
\end{subfigure}%

\begin{subfigure}{.75\textwidth}
  \centering
  \includegraphics[width=1\linewidth]{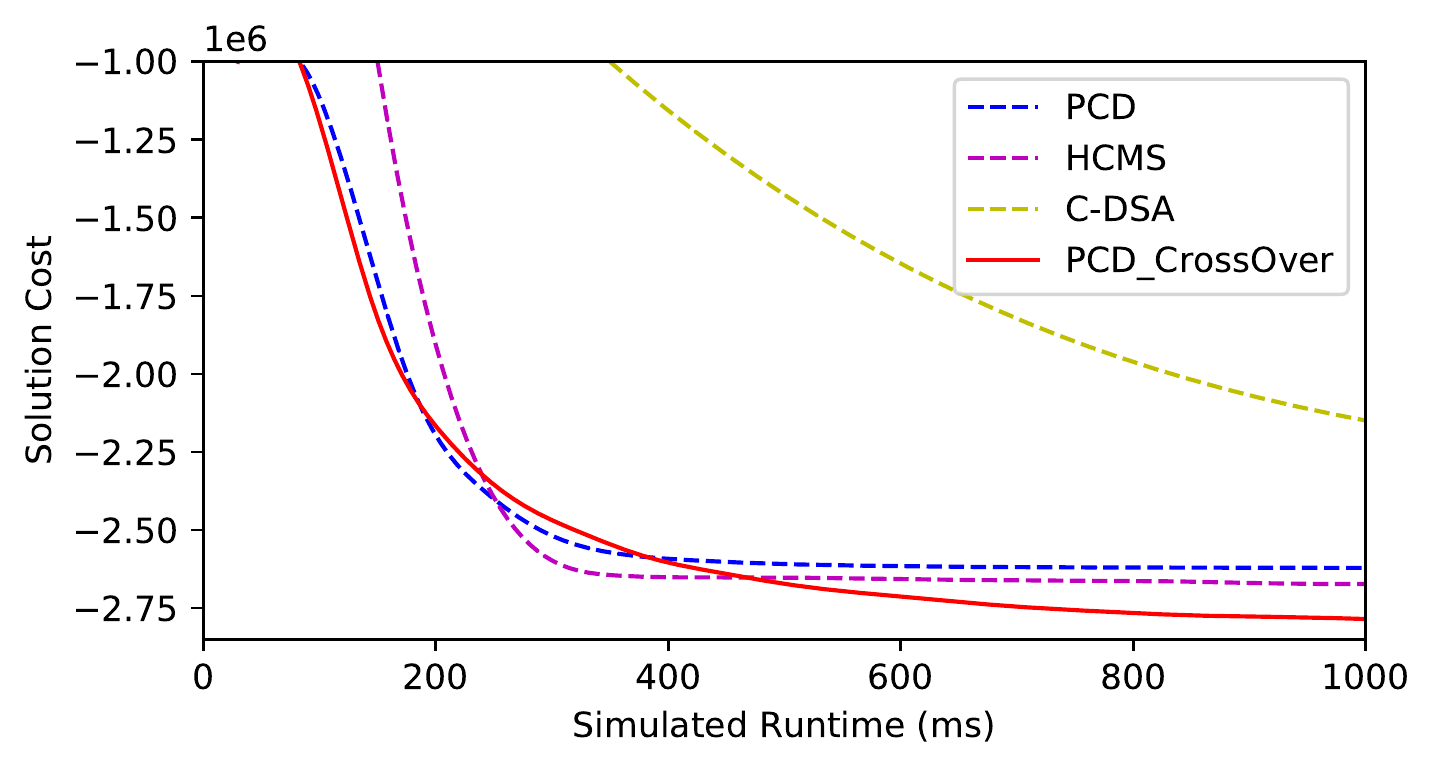}
  \caption{Dense graph}
  \label{fig:dense}
\end{subfigure}
\caption{Solution quality of PCD, PCD\_CrossOver, HCMS, and C-DSA on random graphs.}
\label{fig:random_graph}
\end{figure}

\noindent
In this section, we empirically compare PCD and PCD\_CrossOver to existing state-of-the-art C-DCOP algorithms -- HCMS~\cite{voice2010hybrid} as well as C-DSA and AC-DPOP~\cite{hoang2020new} on the four benchmark problems described in Section~\ref{sec:exp_config}. We follow the literature~\cite{hoang2020new} to determine the parameter values of HCMS, AC-DPOP, and DSA. Specifically, HCMS and AC-DPOP both maintain 3 discrete points per variable, where each discrete point is chosen randomly from the domain range. AC-DPOP moves each of its points $20$ times, where each move is executed by solving a set of gradient equations; and C-DSA uses DSA-B with $p = 0.6$. We set the learning rate of HCMS to 0.01, which is the best value found in our experiments. Finally, it is worth noting that all the differences in the quality of solutions found by the algorithms that we highlight below are statistically significant with $p$-values that are less than $0.01$.

Figure~\ref{fig:random_graph} shows the results of our PCD and PCD\_CrossOver algorithms as well as the existing HCMS and C-DSA algorithms on random graphs. We omit the results of AC-DPOP because it ran out of memory in this setting. In both sparse and dense random graphs, PCD\_CrossOver converges to a better solution compared to all the existing algorithms including PCD. Specifically, PCD\_CrossOver improves PCD by $11.7\%$ in sparse graphs and $10.4\%$ in dense graphs, thereby demonstrating the significance of the crossover procedure described in Section~\ref{sec:crossover}. Moreover, PCD\_CrossOver finds solutions that are better than existing algorithms by about $3.0\% - 13.1\%$ on sparse graphs and $7.6\% - 24.1\%$ on dense graphs after one second. 

\begin{figure}[t]
\centering
  \includegraphics[scale = 0.75]{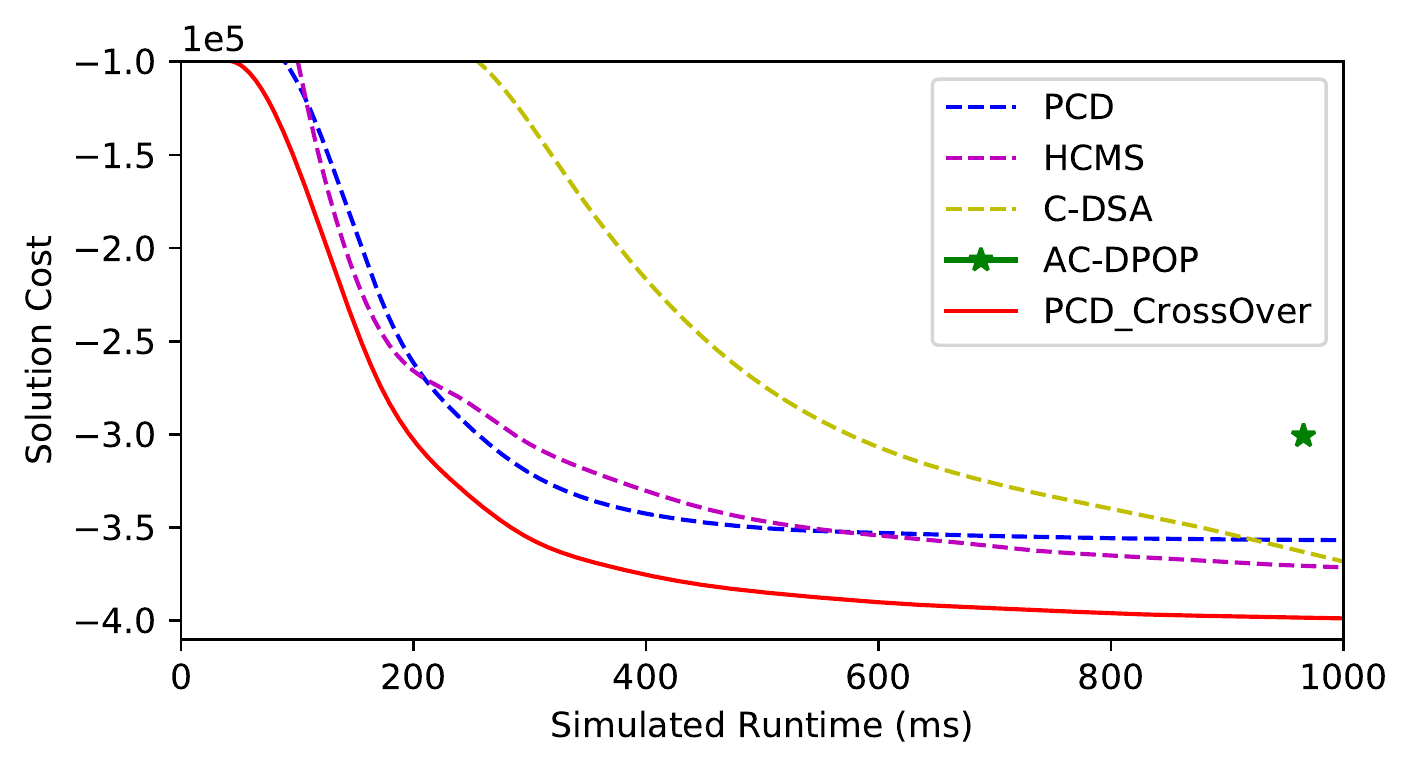}
\caption{Solution quality of PCD, PCD\_CrossOver, HCMS, C-DSA, and AC-DPOP on random trees.}
  \label{fig:tree}
\end{figure}

Table~\ref{tab:agent_table} shows further comparisons on random graph settings varying the number of agents. For this experiment, we run each algorithm for $1500$ms using $30$, $50$, and $70$ agents, each with both sparse (edge probability 0.2) and dense (edge probability 0.6) settings. For smaller graphs (i.e., $|A|=30$), in both sparse and dense settings, the closest competitor of PCD\_CrossOver is C-DSA. For larger graphs and sparse settings, the trends remain the same as the smaller instances. However, as the density and graph size increases, C-DSA takes more time than the competing algorithms and HCMS becomes the closest competitor of PCD\_CrossOver. In all the settings, PCD\_CrossOver outperforms the existing algorithms given the same time. We omit the result for C-DSA in $|A|=70$, $p=0.6$, as it does not produce any output within the given time. A key insight we can draw from this experiment is that neither the graph size nor the density has any adverse effect on the performance of PCD\_CrossOver in random graph settings.

\begin{table}[H]
\centering
\caption{Solution quality of PCD, PCD\_CrossOver, HCMS, and C-DSA on random graphs with different number of agents.}
\label{tab:agent_table}
\resizebox{\textwidth}{!}{\begin{tabular}{@{}cccccc@{}}
\toprule
 &  & PCD\_CrossOver & PCD & HCMS & C-DSA \\ \midrule
\multirow{2}{*}{$|A|$ = 30} & $p$ = 0.2 & -504,082 & -469,093 & -423,383 & -493,730 \\
 & $p$ = 0.6 & -1,166,414 & -1,042,611 & -1,055,878 & -1,096,852 \\ \midrule
\multirow{2}{*}{$|A|$ = 50} & $p$ = 0.2 & -1,151,581 & -1,030,328 & -974,416 & -1,118,344 \\
 & $p$ = 0.6 & -2,897,026 & -2,623,534 & -2,730,943 & -2,334,385 \\ \midrule
\multirow{2}{*}{$|A|$ = 70} & $p$ = 0.2 & -2,129,397 & -1,858,646 & -2,049,757 & -1,636,508 \\
 & $p$ = 0.6 & -5,157,537 & -4,494,102 & -5,060,671 & --- \\ \bottomrule
\end{tabular}}
\end{table}

Figure~\ref{fig:tree} shows the results of our PCD and PCD\_CrossOver algorithms as well as the existing HCMS, C-DSA, and AC-DPOP algorithms on random trees. The observations and trends from the experiments on random graphs are comparable to random trees as well, where PCD\_CrossOver converges to the best solution compared to all other algorithms. Specifically, PCD\_CrossOver improves PCD by $16.2\%$ and finds solutions that are better than existing algorithms by about $6.3\% - 8.3\%$ after one second. 
\begin{figure}[t]
\centering
  \includegraphics[scale = 0.75]{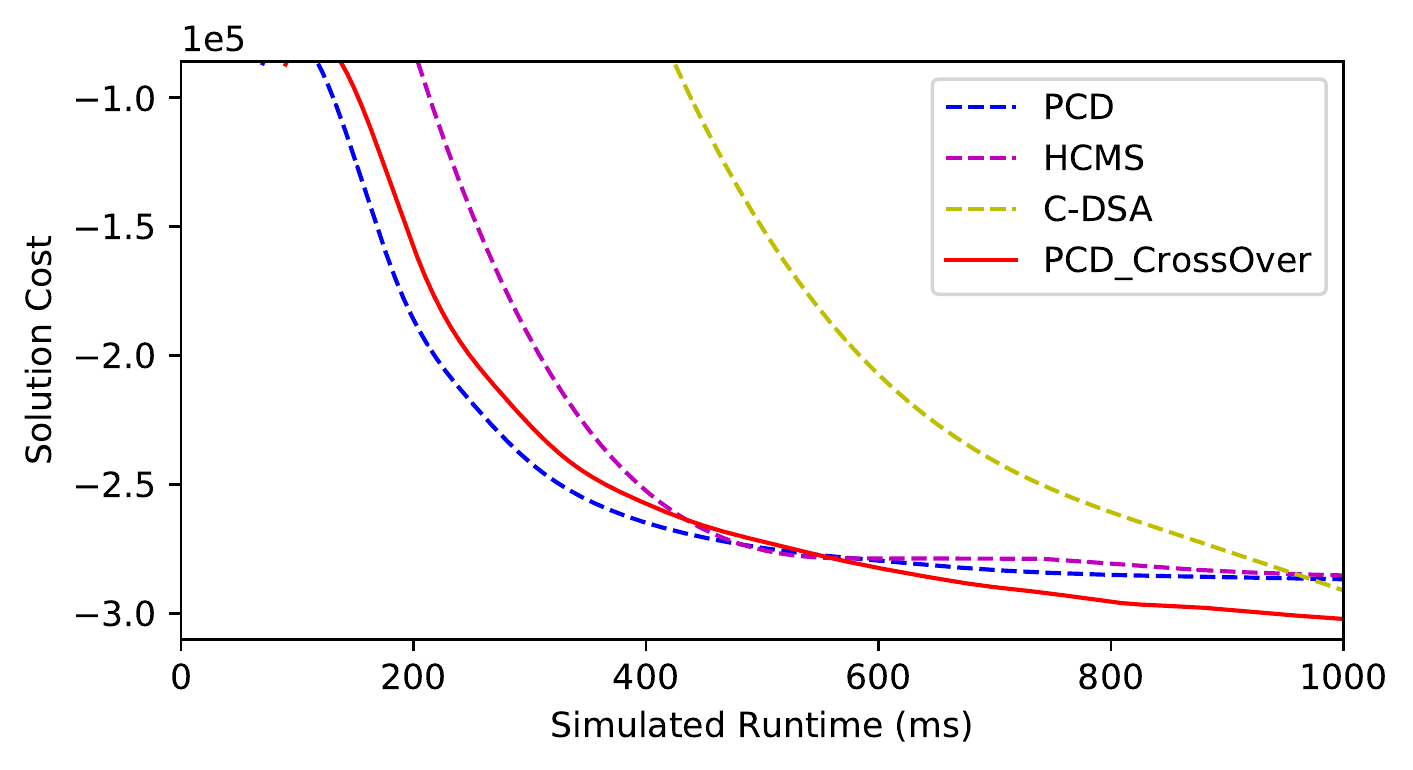}
\caption{Solution quality of PCD, PCD\_CrossOver, HCMS, and C-DSA on Scale-Free Graph.}
  \label{fig:scalefree}
\end{figure}
Figure~\ref{fig:scalefree} shows the performance comparison for scale-free networks in a larger graph setting ($|A| = 100$). Although, PCD, C-DSA, and HCMS show similar performances, PCD\_CrossOver outperforms the existing algorithms by a margin of $22.28\%$ for HCMS and $14.56\%$ for C-DSA.  

\begin{figure}[t]
\centering
  \includegraphics[scale = 0.75]{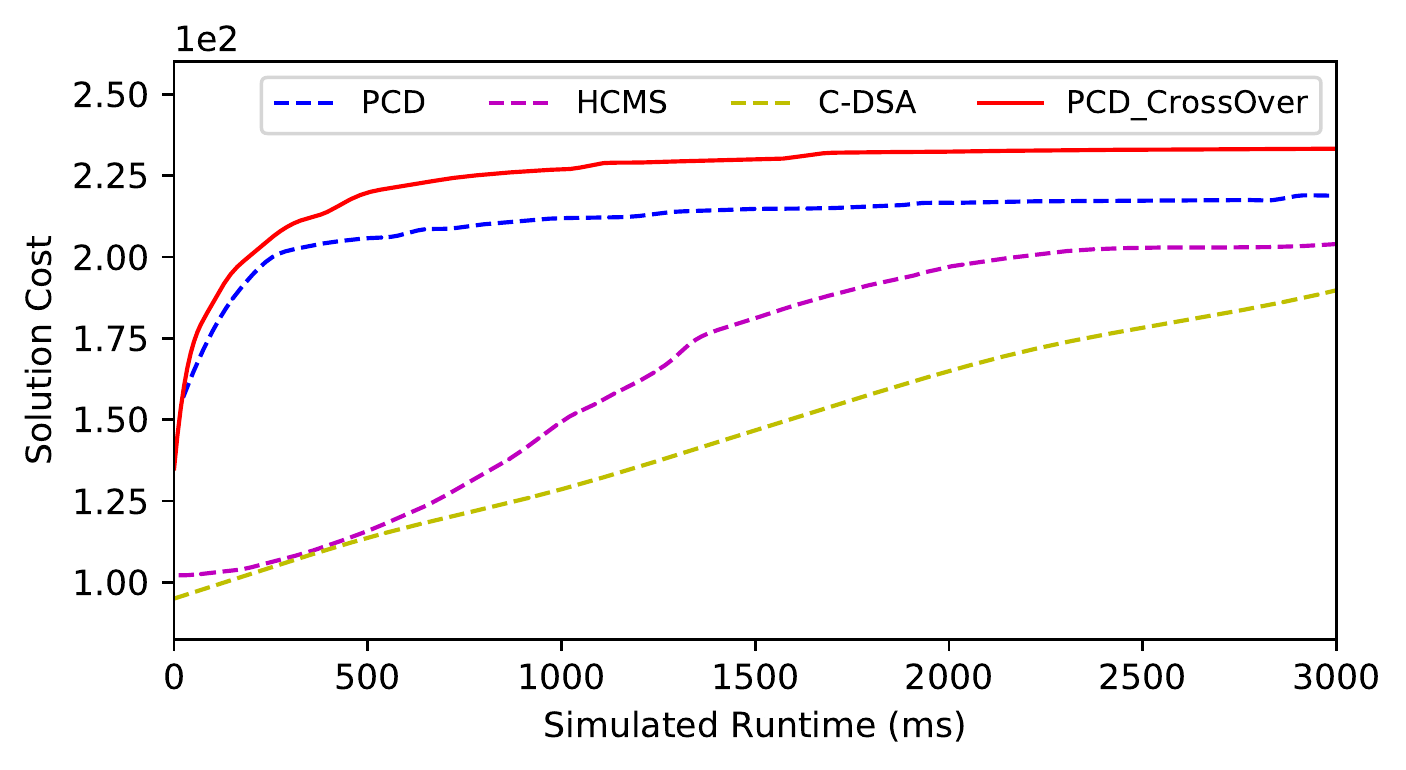}
\caption{Solution quality of PCD, PCD\_CrossOver, HCMS, and C-DSA on random sensor network problems.}
\label{fig:sensor}
\end{figure}

Finally, Figure~\ref{fig:sensor} shows the results of our PCD and PCD\_CrossOver algorithms as well as the existing HCMS and C-DSA algorithms on random sensor network problems. Similar to random graphs, we omit the results of AC-DPOP because it also ran out of memory in this setting. However, unlike the previous two problem settings, we change the optimization problem from one that minimizes the cost of solutions to one that \emph{maximizes} the cost of solutions to stay consistent with actual sensor networks, where the goal is to maximize signal strengths. The general trends from the previous two problem settings are also applicable here. To be exact, PCD\_CrossOver also converges to the best solution compared to all other algorithms. However, interestingly, PCD also finds better solutions than HCMS and C-DSA in this problem setting. In contrast, they all converge to solutions of similar quality in the previous two problem settings. Specifically, PCD\_CrossOver improves PCD by $6.1\%$, and both PCD\_CrossOver and PCD outperform HCMS and C-DSA by about $12.6\% - 17.3\%$ and $6.9\% - 11.9\%$, respectively.

Therefore, the results in these four problem settings clearly demonstrate that PCD\_CrossOver finds better solutions than existing state-of-the-art C-DCOP algorithms, highlighting the promise of particle swarm optimization based approaches to solve continuous C-DCOPs.

\section{Conclusions and Future Work}\label{section:future}
\noindent
\emph{Distributed Constraint Optimization Problems} (DCOPs) have been used to model a number of multi-agent coordination problems. However, its use of discrete variables prevents it from accurately modeling problems with continuous variables. To overcome this limitation, researchers have proposed \emph{Continuous DCOPs} (C-DCOPs), where the key change is that the variables are now continuous instead of discrete, as well as a number of algorithms to solve them. However, existing C-DCOP algorithms primarily rely on gradient-based optimization methods that require derivative calculations. Consequently, they are not suitable for non-differentiable optimization problems. 

To remedy this limitation, we proposed a new approach that generalizes the centralized particle swarm optimization algorithm to a decentralized setting. The new algorithm, called PCD, and its variant that uses crossover operations, called PCD\_CrossOver, maintains a set of particles in a decentralized manner, where each particle represents a candidate solution. They iteratively ``move'' the particles using a series of update equations, which corresponds to updating the solutions maintained over time. Upon termination, the algorithms return the best position over all particles and time steps, which corresponds to the best solution found. We provide theoretical proof for the anytime behavior of our algorithms and show empirical evidence that it outperforms existing state-of-the-art C-DCOP algorithms on four different benchmarks. 

In the future, we plan to further investigate the potential of other population-based algorithms, such as \emph{Artificial Bee Colony} (ABC)~\cite{karaboga2007powerful} and \emph{Cuckoo Search} (CS)~\cite{yang2009cuckoo}, to solve DCOPs and C-DCOPs. Furthermore, we want to study different variants of PSO (e.g., Cooperative PSO~\cite{van2004cooperative} and Hybrid PSO~\cite{angeline1998using}) and the effect of applying other genetic operators that have been proposed over the last few decades to improve the solution quality of PSO. We are also interested in exploring whether our algorithms can be generalized to solve multi-objective C-DCOPs, which, to the best of our knowledge, has not yet been explored.






\bibliographystyle{elsarticle-num-names}
\bibliography{sample.bib}







\end{document}